\documentclass{ar-astro2e} 
\usepackage{graphicx}
\usepackage{amssymb}
\usepackage{ARAstroBib} 
\begin{document}
%
\newcommand{\araa}{Annu. Rev. Astron. Astrophys.}
\newcommand{\apj}{Ap. J.}
\newcommand{\aj}{Astron. J.}
\newcommand{\apjs}{Ap. J. Suppl.}
\newcommand{\apjl}{Ap. J. L.}
\newcommand{\aap}{Astron. Astrophys.}
\newcommand{\aapr}{Astron. Astrophy. Rev.}
\newcommand{\aaps}{Astron. Astrophy. Suppl.}
\newcommand{\pasp}{Publ. Astron. Soc. Pac.}
\newcommand{\pasj}{Publ. Astron. Soc. Japan}
\newcommand{\physrep}{Physics Reports}
\newcommand{\mnras}{MNRAS}
\newcommand{\nat}{Nature}
\newcommand{\ssr}{Space Science Reviews}
\newcommand{\nar}{New Astronomy Reviews}
\newcommand{\herschel}{\emph{Herschel}}
\newcommand{\spitzer}{\emph{Spitzer}}
\newcommand{\planck}{\emph{Planck}}
\newcommand{\akari}{\emph{Akari}}
\newcommand{\iso}{ISO}
\newcommand{\alma}{ALMA}
\newcommand{\galex}{GALEX}
\newcommand{\wise}{WISE}
\newcommand{\arcmin}{\mbox{$^\prime$}}
\newcommand{\arcsec}{\mbox{$^{\prime \prime}$}}
\newcommand{\lsun}{\mbox{L$_\odot$}}
\newcommand{\msun}{\mbox{M$_\odot$}}
\newcommand{\msunyr}{\mbox{M$_\odot$~yr$^{-1}$}}
\newcommand{\rsun}{\mbox{R$_\odot$}}
\newcommand{\lbol}{$L_{\rm bol}$} 
\newcommand{\lfir}{$L_{\rm FIR}$} 
\newcommand{\lir}{$L_{\rm IR}$} 
\newcommand{\tdust}{$T_{\rm dust}$}
\newcommand{\mstellar}{$M_{\ast}$}     
\newcommand{\ergs}{erg~s$^{-1}$}    
\newcommand{\ergscm}{erg~s$^{-1}$~cm$^{-2}$} 
\newcommand{\mum}{\mbox{$\mu$m}}
\jname{Annu. Rev. Astron. Astrophys.}
\jyear{2014}
\jvol{52}
\ARinfo{--------}

\title{Far-infrared surveys of galaxy evolution}

\markboth{Lutz}{Far-infrared surveys}

\author{Dieter Lutz
\affiliation{Max-Planck-Institut f\"ur extraterrestrische Physik, Postfach 1312, 85741 Garching, Germany email:lutz@mpe.mpg.de}}


\begin{abstract}
Roughly half of the radiation from evolving galaxies in the early universe reaches us in the far-infrared and submillimeter wavelength range. Recent major advances in observing capabilities, in particular the launch of the \emph{Herschel Space Observatory} in 2009, have dramatically enhanced our ability to use this information in the context of multiwavelength studies of galaxy evolution. Near its peak, three quarters of the cosmic infrared  background is now resolved into individually detected sources. The use of far-infrared diagnostics of dust-obscured star formation and of interstellar medium conditions has expanded from rare extreme high-redshift galaxies to more typical main sequence galaxies and hosts of active galactic nuclei, out to $z \gtrsim 2$. These studies shed light on the evolving role of steady equilibrium processes and of brief starbursts, at and since the peak of cosmic star formation and black hole accretion.  This review presents a selection of recent far-infrared studies of galaxy evolution, with an emphasis on \herschel\ results. 
\end{abstract}

\maketitle

\section{The road to \herschel\ surveys}
\label{sect:roadtoherschel}

The very first steps to use far-infrared emission as a tool to study galaxy evolution date back to the pioneering IRAS mission, when 60~$\mu$m source counts in the ecliptic pole region were found to exceed no-evolution models \citep{hacking87}. The \emph{Infrared Space Observatory} (ISO) obtained the first deep surveys at both mid- and far-infrared wavelengths, detecting strong evolution of the luminosity function out to $z\sim 1$ and supporting by plausible extrapolation that such dusty galaxies constitute the cosmic infrared background. ISO also pioneered the application of the rich mid-infrared spectra as a tool of studying energy sources and physical conditions in dusty galaxies \citep[see][for a review]{genzel00}. The \emph{Spitzer Space Telescope} revolutionized mid-infrared surveys and opened the window to direct mid-infrared spectroscopy of faint high-z galaxies \citep[review by][]{soifer08}. The \akari\ mission \citep{murakami07} provided uniquely detailed mid-infrared photometric coverage. Independently and at a similar time, ground-based (sub)millimeter surveys at wavelengths 850~\mum\ -- 1.2~mm detected luminous `SCUBA galaxies' \citep{blain02}, which are among the most actively star forming systems at redshifts $z \sim 2$. 

For most redshifts of interest, both spaceborne mid-infrared and groundbased submillimeter surveys miss the rest frame far-infrared peak that dominates the spectral energy distributions of most galaxies. Extrapolation via template spectral energy distributions (SEDs) is then needed to characterize the energy budget of samples that were selected at these wavelengths. Also, such samples can be significantly biased compared to a truly calorimetric selection by infrared luminosity. When aiming for the crucial deep far-infrared surveys, fully cryogenic space missions with small 60--80~cm diameter primary mirrors, such as \iso, \spitzer, and \akari\/, were strongly limited by source confusion. At 250--500~\mum\/, results from the 2~m BLAST balloon telescope \citep{pascale08} provided a first deep glimpse at the extragalactic sky, but are now superseded by \herschel\ data. The European Space Agency's \emph{Herschel Space Observatory} \citep{pilbratt10}, in operation 2009--2013, for the first time met the need for a combination of sensitivity and large aperture (3.5~m) over the full far-infrared and submillimeter wavelength range, substantially reducing confusion levels. At 350~\mum\ and longer wavelengths, low spatial resolution \emph{Planck} all-sky maps and catalogs \citep{planck13a} support studies of galaxy evolution.

Much of this review is based on \herschel\ photometric surveys and pointed observations using the camera modes of the PACS \citep[70, 100, 160~\mum,][]{poglitsch10} and SPIRE \citep[250, 350, 500~\mum,][]{griffin10} instruments.
We will refer to this range as the `far-infrared', leaving the `(sub)millimeter' terminology to ground based surveys at typically 850\,$\mu$m or longer wavelengths.
Similarly, `Submillimeter galaxy (SMG)' will refer to the type of galaxy detected in these surveys, while equivalent \herschel\ selected sources will be called `dusty star forming galaxy (DSFG)' where appropriate.

Traditionally, studies in the local universe make use of a terminology of `luminous infrared galaxies (LIRGs)' defined by their total 8--1000~\mum\ infrared (IR) luminosity $L_{\rm IR}>10^{11}$~\lsun, and their `ultra-' and `hyper-' luminous ULIRG and HYLIRG equivalents above 10$^{12}$ and 10$^{13}$~\lsun, respectively \citep[e.g.][]{sanders96}. These are handy acronyms, but for the purpose of galaxy evolution studies it is important to recall that connotations of these classifications that were carefully calibrated in the local universe may not apply at high redshift. For example, local ULIRGs are found to be major mergers with unusually dense and warm  interstellar medium. The same cannot necessarily be assumed for their higher redshift equivalents at same infrared luminosity. Where used in this review, the (U)LIRG acronyms should be seen as pure infrared luminosity classifications, without further implications for the nature of the galaxy under study.

\subsection{An inventory of \herschel\ surveys}
\label{sect:inventory}

\begin{figure}
\center
\includegraphics[width=13.cm]{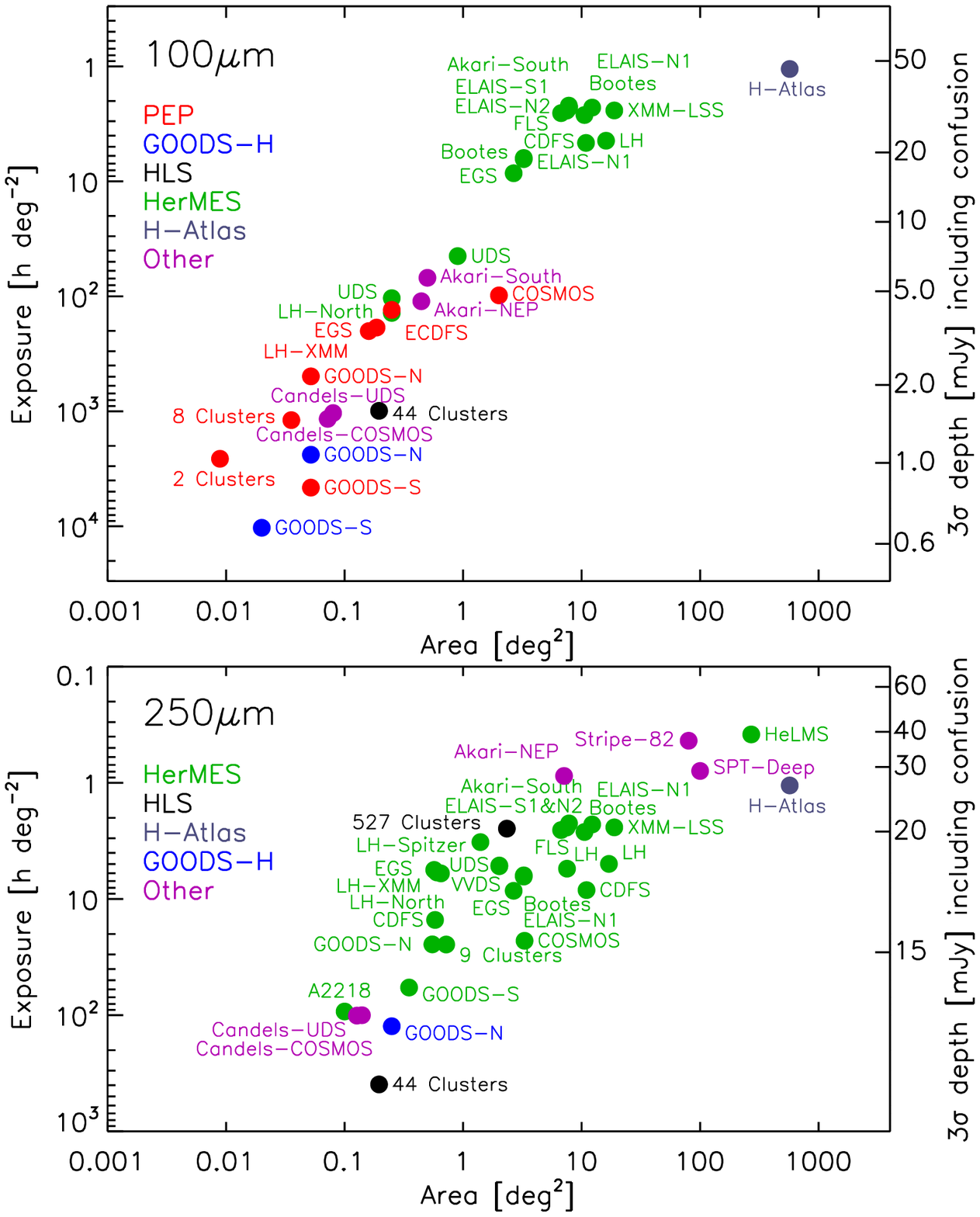}
\caption{Survey area and point source depth reached by some extragalactic surveys with \herschel\ PACS (top, shown for 100~\mum\ but 160~\mum\ data are availabe for the same fields) and SPIRE (bottom, shown for 250~\mum\ but 350 and 500~\mum\ data are available). Exposure is computed from total observing time (including overheads) and survey area. The point source depths (including confusion noise) shown on the right axis should hence be seen as indicative only, since no attempt was made to capture detailed effects of different observing layouts in the various projects. Some sets of cluster observations are included, at the total area summing all objects. Surveys shown are from the projects listed in Section~\ref{sect:inventory} as well as other projects for the Akari-NEP (PI S.~Serjeant), Akari deep field south (PI T.~Takagi), SPT deep field (PI J.~Carlstrom) and SDSS stripe 82 \citep{viero13b}.}
\label{fig:inventory}
\end{figure}

The \herschel\ general purpose observatory served a wide range of science goals, using its capabilities that include both photometric imaging and spectroscopy over the 55--672~$\mu$m wavelength range. Guaranteed and open time for in total about 15\% of \herschel\/'s observing time was devoted to imaging surveys of galaxy evolution, not counting pointed imaging or spectroscopic studies of individual high redshift sources. The list below summarizes some of the major efforts, from the initial guaranteed time and open time \herschel\ key programmes as well as from similar scale programs that were allocated after later calls for proposals. In addition, a good fraction of the science discussed below is based on smaller scale imaging projects as well as pointed imaging and spectroscopic studies which are not listed explicitly. \herschel\ extragalactic surveys exceed 1000 square degrees in total but do not reach the full sky coverage that \planck\ provides at lower spatial resolution for $\lambda\geq 350$~$\mu$m and \akari\ at lower resolution and sensitivity for $\lambda\leq 160$~$\mu$m. Extragalactic confusion limits and instrument design imply a focus for PACS surveys on small area deep surveys, while SPIRE surveys typically emphasize large area. Below we list some major surveys, and Figure~\ref{fig:inventory} gives a graphical overview of area and depth of fields covered.

\begin{itemize}
\item HerMES \citep[][http://hermes.sussex.ac.uk/content/hermes-project]{oliver12} is a wedding cake type multipurpose survey of blank and cluster fields
with focus on SPIRE, covering 70 square degrees in its shallowest tier. A shallow 270 square degree extension has been obtained in the HeLMS project (PI M.~Viero). 
\item PEP \citep[][http://www.mpe.mpg.de/ir/Research/PEP/index.php]{lutz11} is a PACS survey of popular multiwavelength fields such as GOODS, COSMOS, EGS, ECDFS, Lockman-XMM and cluster fields, coordinated with SPIRE observations of the same fields from HerMES. 
\item GOODS-Herschel \citep[][http://hedam.oamp.fr/GOODS-Herschel/]{elbaz11} provides deep PACS and SPIRE observations of GOODS-North and ultradeep PACS coverage of part of GOODS-South. A combined PEP/GOODS-Herschel dataset including all PACS data of the two GOODS fields is published in \citet{magnelli13}.
\item H-ATLAS \citep[][http://www.h-atlas.org]{eales10a} is the largest area shallow extragalactic \herschel\ survey covering about 570 square degrees in SPIRE and PACS.
\item The Herschel Lensing Survey (HLS)\\ \citep[][http://herschel.as.arizona.edu/hls/hls.html]{egami10} provides deep PACS and SPIRE data of 44 X-ray luminous clusters as well as SPIRE snapshots of another 527 clusters. HLS aims at both lensed background objects and at cluster members. 
\item H-CANDELs (PI M.~Dickinson) provides deep \herschel\ data of the CANDELs\\ (http://candels.ucolick.org/index.html) subregions of the COSMOS and UKIDSS-UDS fields, that were not yet covered at equivalent depth by the projects mentioned above.
\end{itemize}

\begin{figure}
\center
\includegraphics[width=13.cm]{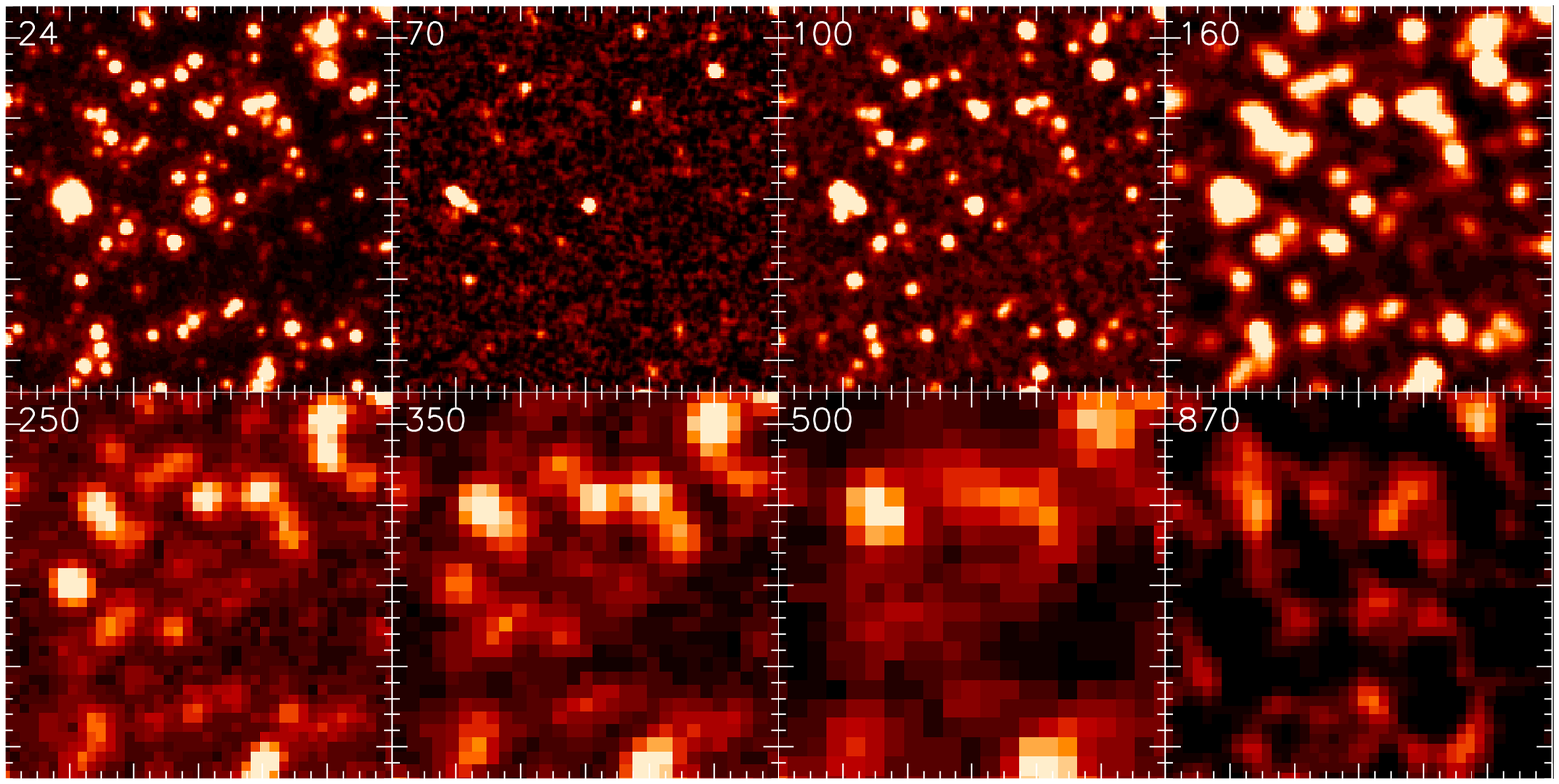}
\caption{Current status of deepest 24--870~$\mu$m infrared surveys, visualized by $4\arcmin\times 4\arcmin$ cutouts in the HUDF region. Data are from the GOODS project (24~$\mu$m), PEP and the combined PEP and GOODS-Herschel data (70--160~$\mu$m), HerMES (250--500~$\mu$m), and the groundbased LESS survey \citep[870~$\mu$m,][]{weiss09}.}
\label{fig:hudf_allwave}
\end{figure}

Figure~\ref{fig:hudf_allwave} visualizes an example for the post-\herschel\ status of deep surveys over the full mid-infrared to submillimeter wavelength range. Surveys with beams of width 5\arcsec\ to 30\arcsec\ are now available over the full range. In the deepest fields, they reach the confusion limit for all wavelengths except 70\,$\mu$m, where a fully cryogenic 3\,m class telescope such as the SPICA project will be needed.

\section{A resolved view of the cosmic far-infrared background}

The detection of an isotropic cosmic far-infrared background (CIB) by the COBE satellite \citep[see][for reviews]{hauser01,lagache05} was immediately interpreted in terms of radiation from galaxies in the early universe, absorbed and re-emitted by their interstellar dust. Strong observational support was provided by retrieving the major part of the CIB when stacking \spitzer\ 70 and 160~$\mu$m data at the positions of mid-infrared detected galaxies \citep{dole06}, and by resolving the majority of its 850~$\mu$m long wavelength tail using the deepest SCUBA counts, assisted by gravitationally lensing clusters of galaxies \citep[e.g.][]{knudsen08,zemcov10a}. Closer to the peak of the CIB, source confusion restricted the smaller cryogenic far-infrared telescopes \iso\ and \spitzer\ to mostly statistical methods, a limitation that is significantly relaxed by \herschel\/'s larger primary mirror and improved wavelength coverage.

\subsection{Number counts and resolved fraction of the far-infrared background}
\label{sect:counts}

\begin{figure}
\center
\includegraphics[width=10.cm]{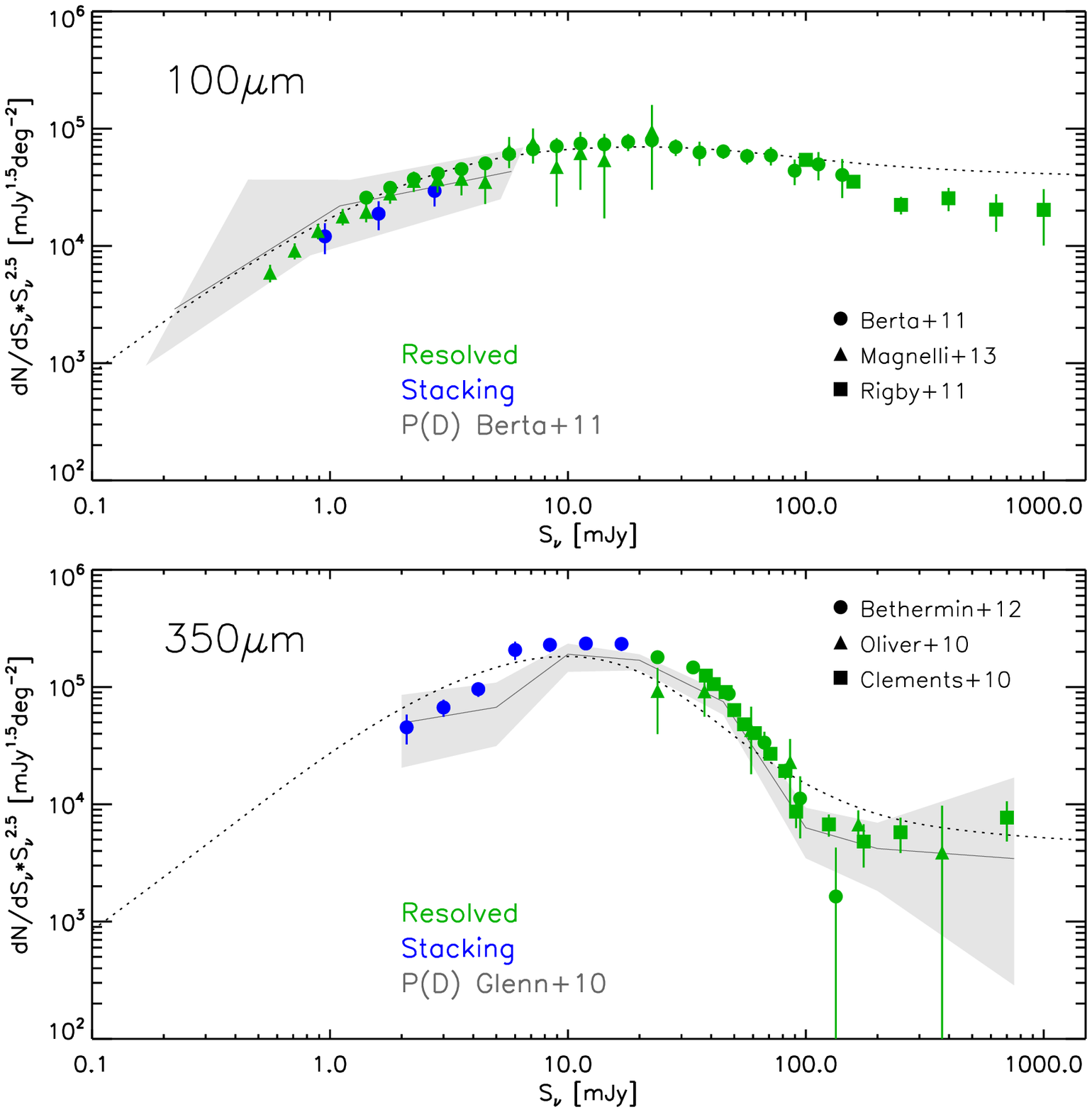}
\caption{Euclidean normalized differential number counts for 100~\mum\ (\herschel\/-PACS) and 350~\mum\ (SPIRE). The references quoted in the figure also provide counts at other \herschel\ wavelengths. The dotted line shows an example for a number count model \citep{bethermin12b}.}
\label{fig:counts}
\end{figure}

The enhanced \herschel\ capabilities were rapidly used for number count studies based on different analysis techniques: Direct resolved detection, stacking analysis, and probability of deflection analysis (P(D)). At 70--160~\mum, \citet{berta10,berta11} and \citet{magnelli13} provide counts of increasing depth from individually resolved detections, based on a variety of deep blank fields, and consistent with the analysis of the lensing cluster Abell~2218 by \citet{altieri10}. \citet{rigby11} extend the 100 and 160~$\mu$m counts to larger area, and \citet{sibthorpe13} reconfirm that cosmic variance is a minor concern for these results. Stacking and P(D) analysis at these wavelengths has been reported by \citet{berta11} but provides only a modest extension compared to the very deep resolved PACS counts. At 250--500~$\mu$m, \citet{oliver10}, \citet{clements10}, and \citet{bethermin12a} provide resolved counts. Because of the higher SPIRE confusion limits, statistical methods are more important and have been applied by \citet[][P(D)]{glenn10} and  \citet[][stacking]{bethermin12a}. At all \herschel\ wavelengths, measurements now reach below the knee in the integral number counts (i.e. the peak in the euclidean normalized differential number counts). An unusual feature of the bright end of the counts at long \herschel\ wavelengths is the significant contribution by gravitationally lensed sources (Section~\ref{sect:lensed}).

Figure~\ref{fig:counts} summarizes number count results for the central wavelengths of the two \herschel\ cameras, again highlighting the importance of statistical methods at the longer wavelengths because of increasing source confusion. Different measures of source confusion are in use and applicable for different shapes of the number counts and location of the detection limit in the counts \citep[see, e.g., discussion in][]{dole04}. For 500, 350, and 250~$\mu$m \citet{nguyen10} report a confusion noise of 6.8, 6.3, and 5.8~mJy. For 160 and 100~\mum\ confusion noise has been measured by \citet{berta11} and using the deepest PACS data by \citet{magnelli13}, who find 0.68 and 0.15~mJy, respectively. No \herschel\ deep survey reaches the extragalactic confusion limit at 70~\mum\ for a 3.5~m telescope \citep{berta11}. 

\begin{figure}
\center
\includegraphics[width=12.cm]{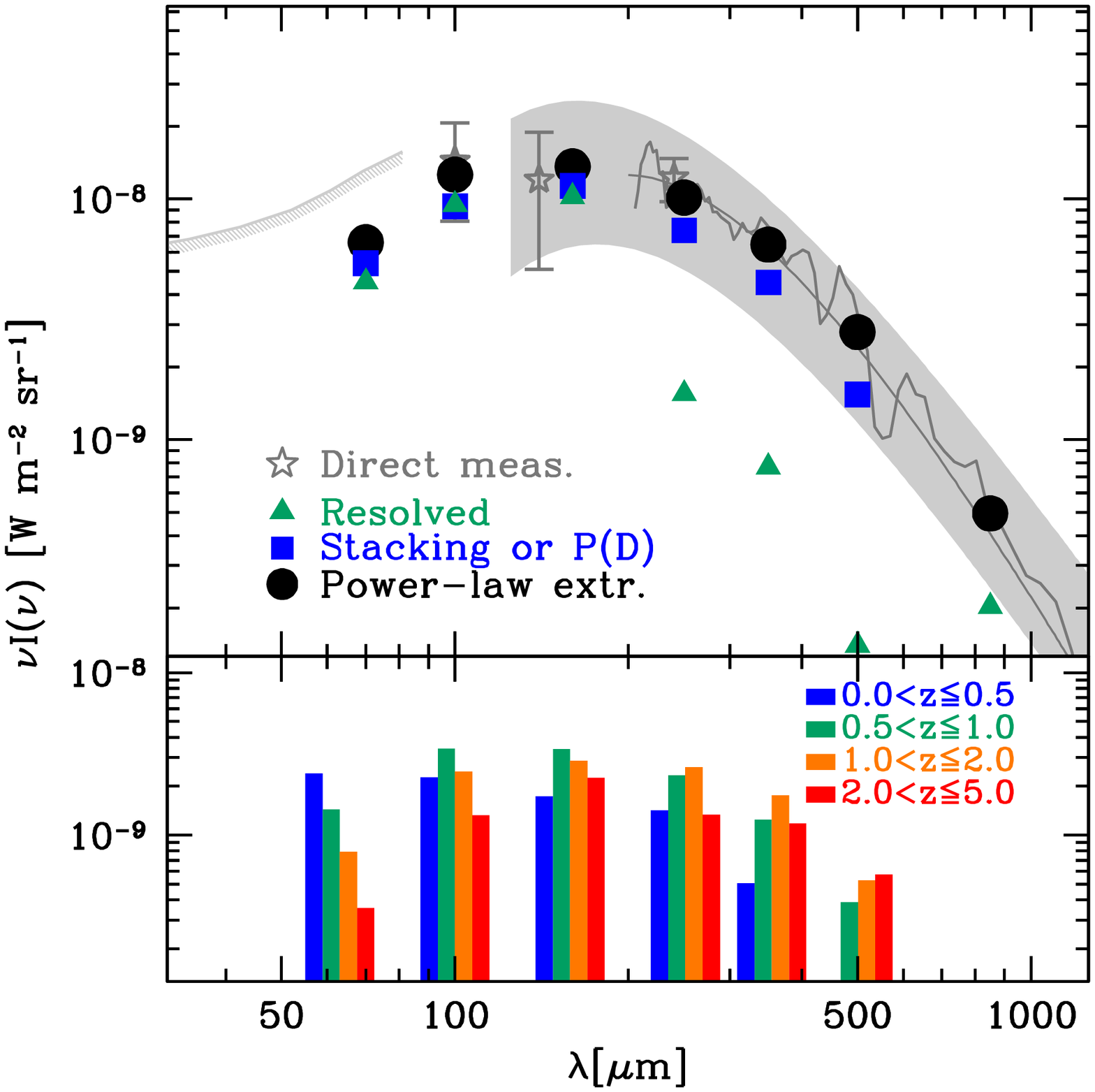}
\caption{The cosmic far-infrared background as seen by direct measurements and as resolved by \herschel\/. Direct measurements (grey) include the $\gamma$-ray based limits of \citet{mazin07}, COBE-DIRBE results as presented in \citet{dole06} (asterisks), the COBE-FIRAS $\lambda >200 \mu$m spectrum of \citet{lagache99} and the modified blackbody fit of \citet{fixsen98}. CIB contributions by resolved sources are from \citet{berta11} and \citet{magnelli13} (70-160~\mum), \citet{bethermin12a} (200--500~\mum) and \citet{zemcov10a} (850~\mum). Stacking results are from \citet{bethermin10} (70~\mum) \citet{berta11} (100, 160~\mum) and \citet{bethermin12a} (250--500~\mum), power law count extrapolations from the same works and \citet{zemcov10a} (850~\mum). The lower panel shows the contributions of different redshift slices to the part of the CIB that is contained in resolved sources (PACS) and covered by stacking (SPIRE) \citep{berta11,magnelli13,bethermin12a}.}
\label{fig:cib}
\end{figure}

In addition to the star formation rates and black hole accretion rates of galaxies, direct predictions of far-infrared counts from semianalytic or hydrodynamical models of galaxy evolution are sensitive to the amount and geometry of obscuring dust. Amount of dust and in particular its poorly predictable distribution affect energy and shape of the far-infrared SED. For that reason among others, infrared counts have been traditionally interpreted in terms of backward evolution models which adopt locally calibrated empirical SEDs or families of SEDs. These models then predict counts on the basis of an assumed evolution of luminosity and number densities of the luminosity functions for the populations that are represented by these SEDs. Already before the launch of \herschel\ \citep[e.g.][]{valiante09}, assumptions on evolution of SEDs at given infrared luminosity started to be included as well. As noted in many of the works presenting observed \herschel\ counts, pre-\herschel\ models often gave decent representations of the observed counts in some individual bands, but typically did less well in providing a consistent picture for all available far-infrared bands, and in particular for the redshift distributions. It is inevitable that future more successful models include the \herschel\ evidence on evolution of SEDs (see Section~\ref{sect:sed}), as already attempted in the model of \citet{bethermin12b} which implements SEDs for galaxies on the evolving main sequence of star forming galaxies and for starbursts above this sequence. The same constraints on adopted SEDs as for the simple backward models will be faced by the more physical forward semianalytic or hydrodynamical models.

Figure~\ref{fig:cib} summarizes the 70--850~$\mu$m cosmic far-infrared background as
directly measured by COBE, and the contributions of galaxies as determined by \spitzer\/, \herschel\/, and groundbased submillimeter surveys. The extrapolation of measured number counts is consistent with the direct CIB measurements, and provides smaller uncertainties at some wavelengths near the peak of the CIB. Combining extrapolated \spitzer\/, \herschel\/, and 850\,$\mu$m results, the total 8--1000~\mum\ background is $27^{+7}_{-3}$~nW~m$^{-2}$~sr$^{-1}$ \citep{bethermin12a}. There is no evidence for a diffuse extragalactic IR background that is not accounted for by galaxies. The CIB fraction accounted for by individually resolved galaxies peaks at 75\%\ at 100 and 160~$\mu$m \citep{magnelli13}. The resolved fraction is lower at 70~$\mu$m where \herschel\ could not integrate to its confusion limit \citep{berta11}, and at 250--500~\mum\ with 15--6\%\ of the CIB resolved at these wavelengths due to increasing source confusion \citep{oliver10,bethermin12a}. At 250--500~\mum, stacking analysis again attributes 73 to 59\% of the CIB to known galaxies \citep{bethermin12a}. As expected, the typical redshift of sources contributing to the CIB increases with wavelength (lower panel of Figure~\ref{fig:cib}). Of the major CIB fraction that is directly resolved or retrieved by stacking, half of the emission at wavelengths 100, 160, 250, 350, and 500~\mum\ is contributed by sources above $z=0.75$, 1.00, 1.04, 1.20, and 1.25 \citep{magnelli13,bethermin12a}. Median redshifts of sources with fluxes above faint limits are naturally somewhat higher (e.g. $z_{\rm Med}=1.37$ above $S_{100}=1.5$~mJy, $z_{\rm Med}=1.22$ above $S_{160}=2.5$~mJy).

\subsection{Extinction insensitive star formation rates and comparison to other tracers}
\label{sect:recalibsfr}

Far-infrared emission is an established tool to determine star formation rates of galaxies. The total infrared luminosity provides an observationally straightforward calorimetric measure that can be interpreted as total star formation rate (SFR), if considering a number of aspects. 

First, the infrared traces only the obscured part of the radiation emitted by the ensemble of young stars in a galaxy, the part which is absorbed by dust. A short wavelength tracer has to be used to quantify the escaping fraction, in order to either ensure dominance of the obscured star formation, or revert to a composite star formation indicator that adds the obscured and unobscured star formation. In massive high-z star forming galaxies, dust obscured star formation is typically about 5--10 times stronger than unobscured star formation (Section~\ref{sect:lbglae}).

Second, the contribution to the IR luminosity by accretion onto an active galactic nucleus (AGN) has to be low in order for star-formation to dominate. 
This is discussed in Section~\ref{sect:agnhosts}, but in short the generally elevated star formation in high redshift galaxies implies that for moderate luminosity AGN from typical surveys, host stellar heated emission will dominate over AGN heated emission in the far- and sometimes even mid-infrared. Elimination of known AGN on the basis of X-ray data and of rest frame near-infrared hot dust `power laws' can further reduce this source of uncertainty.

Third, dust can also be heated by older stars, which can dominate the dust heating in regions such as the bulges of quiescent local spirals. This makes the conversion from total infrared (8--1000\,$\mu$m) luminosity to current star formation rate more dependent on the star formation history than for tracers of O stars, such as the H$\alpha$ line which originates only in the ionized gas. Compared to the assumption of constant star formation over 100~Myr, as in the widely adopted conversion of \citet{kennicutt98} \citep[note update in][]{kennicutt12}, constant star formation over 10~Gyr will increase the ratio of IR luminosity and current SFR by almost a factor 2, an effect that can increase further for decaying star formation histories.
Further aspects of multiwavelength methods to determine star formation rates, including updated calibrations, are covered in the recent review by \citet{kennicutt12}.

\subsubsection{Rest frame far-infrared star formation rates}

Deriving total 8--1000~$\mu$m IR luminosities and star formation rates by fitting SED templates is straightforward for fully sampled \herschel\ SEDs, but sensitivity and confusion limits often limit detections to one or two photometric bands only. Several studies confirm that, as long as one of these detected photometric points is near the far-infrared SED peak, such monochromatic based IR luminosities are good to $\pm$0.1--0.2~dex even if considering a wide range of possible SED shapes \citep{elbaz11,nordon12,berta13a,magnelli13}. For the fitting, SEDs derived from or validated on \herschel\ data should be used, since lack of observational constraints caused some earlier SED families to show too low fluxes on the long wavelength side of the far-infrared (FIR) peak \citep{elbaz10}.

For the high redshift star forming galaxies that are individually detected in \herschel\ surveys, the ratio of infrared and ultraviolet (UV) luminosity \lir\//$\nu$L$_\nu(UV)$ (also called the infrared excess IRX) is typically of order 10, sometimes higher \citep[e.g.][Section~\ref{sect:lbglae}]{buat10,buat12,nordon13}. These \herschel\ detections are mostly galaxies near and above the `main sequence' of star forming galaxies, the narrow locus of most star forming galaxies in the star formation rate - mass plane \citep[e.g.][see also Section ~\ref{sect:ms}]{brinchmann04,noeske07,elbaz07,daddi07}.
Only rarely do galaxies with individual \herschel\ detections fall short of $IRX\sim 2$ \citep[e.g., in the UV selected sample of][]{oteo12a}. Converting from the ratio of luminosities to SFR ratios, obscured star formation thus dominates safely and IR based SFRs are useful even without adding the unobscured UV, as is of course still recommended if ultraviolet data are available. More care in this respect is needed for lower star formation rate and lower obscuration samples for which stacked \herschel\ detections can still be obtained \citep[e.g. the samples of][with $IRX\sim 7$ and $\sim 3-4$]{reddy12,lee12}.

Comparing IR based star formation rates with the scarcer H$\alpha$ star formation rates (extinction corrected using the Balmer decrement), studies find good agreement. This applies both to a sample of mostly near main-sequence star forming galaxies with \herschel\ and H$\alpha$ detection at $z\sim 0.27$, $SFR\sim 6$~\msunyr\ \citep{dominguez12}, and a sample of $z\sim 1.4$, $SFR\sim 100$~\msunyr\ SPIRE-selected galaxies near and above the main-sequence \citep{roseboom12a}. The Balmer decrement suggests these samples are moderately dusty (E(B-V)$\sim$0.5). These findings suggests a minor importance of dust heating by old stars in these star forming objects. Also, despite sizeable star formation rates,  these galaxies cannot have a large contribution of heavily obscured star formation that is not traced by H$\alpha$.  More care in interpreting the far-infrared luminosity is needed for passive objects well below the star forming sequence. Given realistic survey depths, this occurs for individual detections mostly for a subset of $z<0.5$ detections, and for stacks of other passive samples. 
 
\subsubsection{Star formation rates from mid-infrared emission}

Given the superb \spitzer\ mapping speed at 24$\mu$m, mid-infrared surveys remain a key resource, including many sources that are too faint for \herschel\ detection. Deducing star formation rates for high-z galaxies from 24$\mu$m fluxes requires an extrapolation to \lir\ that is more uncertain, given the diverse factors that are shaping this spectral range: Continuum from warm or transiently heated dust in HII regions, aromatic `PAH' emission features, possibly AGN heated emission, and silicate absorption.

Traditionally, this extrapolation was done using locally calibrated luminosity dependent spectral templates \citep[e.g.][]{chary01}. These encode the physical properties of local infrared galaxies. Specifically, the ratio of 8$\mu$m PAH emission to total infrared is lower for local ULIRGs than for lower luminosity galaxies. This relates to the compact star forming regions and intense radiation fields of these galaxy mergers ULIRGs. Already during the \spitzer\ era, observations suggested that at $z\sim 2$ application of these templates leads to overpredicted IR luminosities \citep{papovich07,daddi07}. This `mid-IR excess' was ascribed to either relatively stronger PAH emission in $z\sim 2$ galaxies, or to a strong AGN mid-IR continuum.

\herschel\ observations have quantified this mismatch for large samples and for individual detections. There is a factor $\sim 5$ overprediction of $z\sim 2$ SFRs if using 24~\mum\ photometry and typical locally calibrated template families  \citep{nordon10,elbaz10,elbaz11,nordon12}. The effect conspicuously sets in close to $z\sim 2$ when the strongest PAH feature enters the MIPS 24~$\mu$m band \citep[][see also Figure~\ref{fig:ms}]{elbaz10,elbaz11}, suggesting that it is due to enhanced PAH emission rather than due to AGN continuum. This is unambiguously confirmed \citep{nordon12} by fully reproducing the photometric trends by trends in the PAH emission in the ultradeep low resolution \spitzer\ spectra of $z\sim 1$ and $z\sim 2$ galaxies by \citet{fadda10}. Section~\ref{sect:sed} discusses the relation of these findings to the evolving star forming main sequence, and to changes with redshift in the interstellar medium conditions of galaxies of a given infrared luminosity.

Several procedures are in use to avoid the overprediction that arises when applying local luminosity-dependent templates to MIPS data of high-z sources. The main sequence SED template of \citet{elbaz11} encodes the proper 8$\mu$m/IR ratio that is valid for the bulk of MIPS galaxies except the minor fraction of starbursts above the main sequence. Similar arguments apply to the single template of \citet{wuyts08}. \citet{nordon12} provide a procedure to choose the right template as a continuous function of specific star formation rate ($SSFR=SFR/M_\ast$) offset from the main sequence, while \citet{rujopakarn13} derive $z\sim 2$ relations by treating high-z galaxies as main-sequence objects, and linking their SEDs to local objects of same IR surface density. If sufficient UV to mid-IR SED information is available, it is also possible to derive IR luminosities by rescaling templates from libraries that span SED shapes without a fixed link of shape to IR luminosity. These include the library of \citet{polletta07} and the derivation by \citet{berta13a} that also considers \herschel\ data. \citet{nordon12} and \citet{berta13a} compare some of these revised prescriptions. Given that methods which trace the changes of SEDs with SSFR offset from the main sequence will amplify noise in the 24~\mum\ photometry and internal spread in the SEDs at given LIR, there is a tradeoff between simple single templates (providing low scatter, but residual systematics in the comparison of \lir\ predicted from 24~\mum\ data and true \lir), and the larger scatter but lower systematics in the more complex approaches.    

The methods discussed above have been tested using MIPS and \herschel\ detections, i.e. effectively for the main sequence above $\sim 10^{10}$ ($10^{10.5}$) \msun\ at $z\sim 1$ (2), plus more actively star forming galaxies. Application to lower mass and lower SFR main sequence objects may need further validation, note that the stacked $L_{\rm IR}/L_{8\mu{\rm m}}=8.9\pm1.3$ reported by \citet{reddy12} for fainter UV-selected galaxies is about 80\% larger than the main sequence value, possibly due to a weakening of PAH for more compact star formation and/or lower metallicity.

\subsubsection{Radio emission and the radio -- far-infrared correlation}

Thirty years after its discovery, the tight correlation between the radio emission and far-infrared emission of star forming galaxies is still a challenge for models \citep[e.g.][and references therein]{lacki10a,lacki10b}. Empirically, the radio--far-infrared correlation enables interferometric radio continuum surveys to provide a convenient high spatial resolution and extinction insensitive access to high-z star formation rates. This is true as long as no AGN related excess radio emission is present, and changes of the radio--far-infrared correlation with redshift are absent or calibrated. Supplementary radio interferometric data have hence played a key role, for example, in identifying and characterizing the submillimeter galaxy (SMG) population detected in groundbased surveys (see also Section~\ref{sect:smg}). 

Again, verification of the relation at high redshift, but with minimal extrapolation uncertainties, needs samples of star forming galaxies with SED detections near the rest frame far-infrared peak. Selection effects in both radio and far-infrared are of paramount consideration for current studies. First results using early \herschel\ datasets in combination with VLA surveys give no compelling evidence for an evolution of the radio--far-infrared relation out to z$\sim$2 \citep{ivison10,jarvis10} but are also consistent with modest evolution towards a lower ratio of IR to radio emission. Combining full \herschel\ data with upcoming JVLA radio surveys should significantly enhance the power of this approach, and guide the further use of deep radio surveys to measure star formation rates. 

\subsubsection{Comparison to extinction corrected UV emission and SED fitting}
\citet{wuyts11a} compare star formation rates from rest-UV to near-IR SED fitting to total star formation rates that add unobscured SFR$_{\rm UV}$ and re-emitted SFR$_{\rm IR}$ (directly measured by \herschel\ or \herschel\/-recalibrated 24~$\mu$m), using a deep GOODS-S dataset out to z$\sim$3. Star formation rates from SED fitting and total SFRs are reasonably consistent at low and intermediate star formation rates, but SED fitting underpredicts the total SFR$_{\rm UV+IR}$ for galaxies with the largest ratio  $SFR_{IR}/SFR_{UV}$, typically found at the highest $SFR_{UV+IR} >100$~\msunyr\ and at $z\gtrsim 2.5$.

A mixed picture emerges from studies comparing \herschel\ far-infrared star formation rates with star formation rates estimated from UV emission and slope, via the IRX-$\beta$ relation \citep[e.g.][]{meurer99}. While some populations match this relation
on average \citep{reddy12}, the scatter is large (see also Section~\ref{sect:lbglae}). Specifically, the most infrared luminous galaxies often have much larger IRX than corresponding to this relation, reminiscent of the situation for local ULIRGs \citep{goldader02}. This offset correlates with the offset from the star forming main sequence \citep{nordon13}.

\subsubsection{Summary}
Given these constraints, studies of star forming galaxies over a wide range of redshifts and star formation rates may often benefit from a combined approach, which in order of decreasing star formation rate uses:
\begin{itemize}
\item Far-infrared star formation rates for galaxies with at least one \herschel\ detection near the SED peak.
\item Mid-infrared star formation rates, using recipes that have been validated against \herschel\ data.
\item Star formation rates from rest UV to NIR SED fitting.
\end{itemize}
Specific limits between these steps, and the need to invoke all or just part of these steps will also depend on survey practicalities. Examples of studies utilizing such an approach include \citet{wuyts11b}, \citet{rodighiero11}, \citet{wang13}, and \citet{magnelli14}.

\subsection{Evolution of the infrared luminosity function and the star formation rate density}
\label{sect:lf}

Evolution of the infrared luminosity function and the infrared luminosity density encode the cosmic history of dusty star formation, and are an essential part of measuring the total cosmic star formation history. Previous determinations were mostly based on ground-based submillimeter data (requiring significant extrapolation to the total infrared emission, and covering only a small luminosity range) or \spitzer\ mid-infrared data (also requiring significant extrapolation to the total infrared, Section~\ref{sect:recalibsfr}).

\begin{figure}
\center
\includegraphics[width=13.cm]{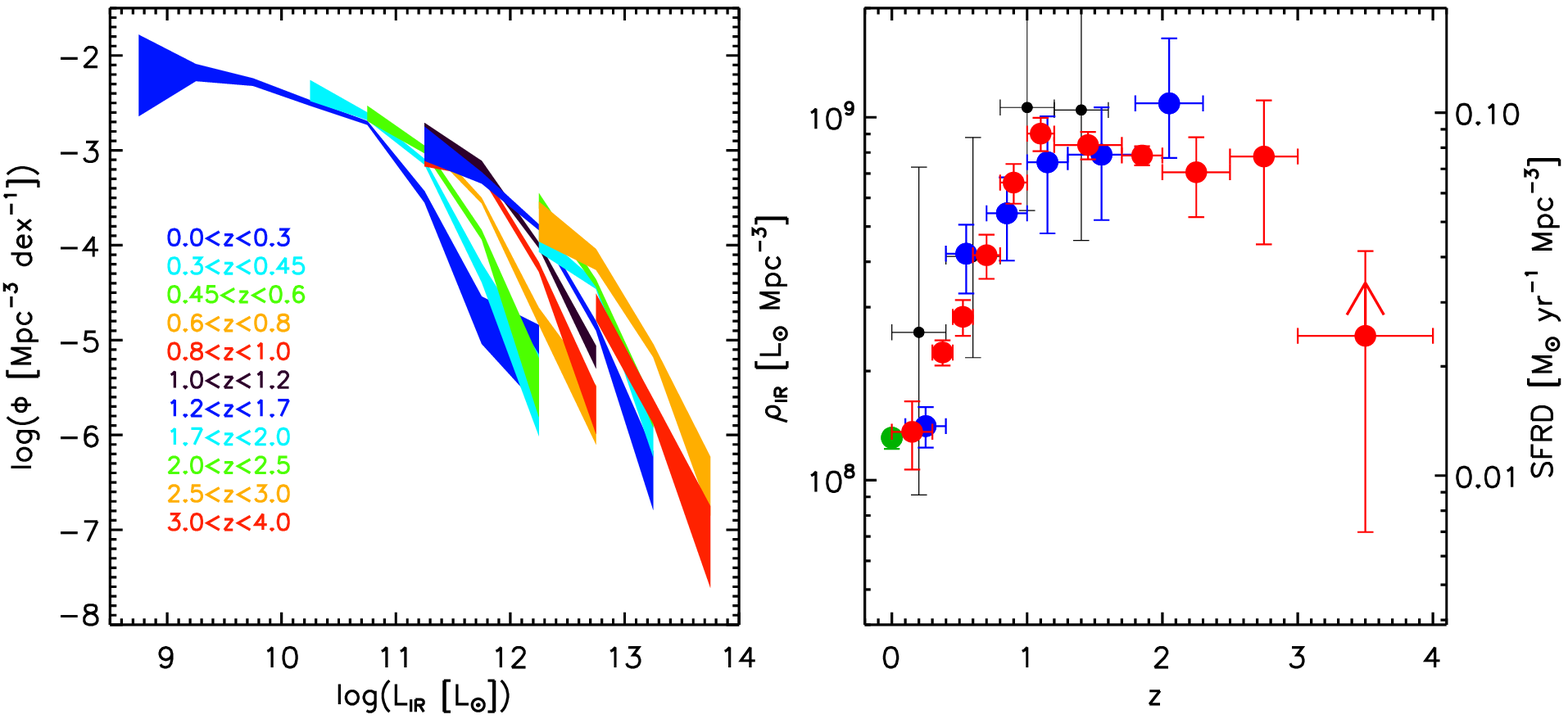}
\caption{{\bf Left}: Evolution of the total infrared (8--1000~$\mu$m) luminosity function. Results are from \citet{gruppioni13}, the width of the band indicating the $\pm 1\sigma$ poissonian error. {\bf Right}: Evolution of the infrared luminosity density. Data are from
\citet[][red]{gruppioni13}, \citet[][blue]{magnelli13}, \citet[][black]{casey12b}, and the local results of \citet[][green]{sanders03}. As argued by \citet{gruppioni13}, the $z\sim 3.5$ point may be a mild lower limit due to selection at $\lambda<200$~$\mu$m. Adopted cosmology and initial mass function are as in \citet{gruppioni13}. The right axis (star formation rate density) represents only the dust-obscured part of star formation.}
\label{fig:lf}
\end{figure}

Early \herschel\ results clearly demonstrate strong evolution of the total and monochromatic infrared luminosity functions \citep[e.g.][]{gruppioni10,vaccari10,dye10,eales10b,lapi11} and, at $z\sim 0$,  dependence on environment with the Virgo cluster showing a peaked luminosity function \citep{davies10,davies12}. Perhaps the most complete analyses of the evolving total infrared luminosity function are provided by \citet{gruppioni13} from a large combined PACS and SPIRE dataset at z$\lesssim$4 and by \citet{magnelli13} at $z \lesssim 2$ using the deepest \herschel\/-PACS data. Figure~\ref{fig:lf} shows evolving total IR luminosity functions and infrared luminosity densities from these and other references. Given depth limits of the surveys used, the faint end slope of the luminosity function is only weakly constrained at all but the lowest redshifts (Figure~\ref{fig:lf}), and assumptions vary. For, example at $1.0<z<1.2$ and in the $L_{\rm IR}<10^{11.2}$~\lsun\ range that is not constrained by \herschel\/, \citet{gruppioni13} adopt a slope of the luminosity function $\Phi={dN/( d(logL_{\rm IR}) dV)}\propto L_{\rm IR}^{-0.2}$ [Mpc$^{-3}$ dex$^{-1}$], guided by their lowest redshift bin, while \citet{magnelli13} adopt $\Phi\propto L_{\rm IR}^{-0.6}$. The deepest \spitzer\ 24~$\mu$m data provide some guidance at these luminosities \citep{magnelli13}, in a regime where extrapolations from mid-infrared to total infrared are rather certain, but cannot fully resolve the ambiguities in slope. The effect of these faint end slope differences on the integrated infrared luminosity density is small, however.

In summary, infrared luminosity functions and infrared luminosity density are now determined by direct rest far-infrared measurements out to $z>3$. The contribution of `LIRG' and `ULIRG' luminosity bins to the luminosity density increases from 5\% and 0.4\%  in the local universe to 31\% and 50\% at $z\sim 2$ \citep{magnelli13}. This does not necessarily imply that these important $z\sim 2$ luminous objects physically resemble local (U)LIRGs. Most of the $z=1-2$ IR luminosity density arises in near main sequence objects rather than starbursting outliers, which dominate only the upper end of the luminosity function \citep{gruppioni13}.
Considering stellar mass bins, the IR luminosity density is dominated by the 
range $10^{10}< M[\msun] <10^{11}$ at all redshifts studied by \citet{gruppioni13}.

\citet{burgarella13} combine the results of \citet{gruppioni13} with recent UV luminosity functions and compare the luminosity densities $LD_{\rm IR}$ and 
$LD_{\rm UV}$ obtained by integrating the respective luminosity functions. The ratio of infrared to UV luminosity densities traces the average obscuration of the galaxy population and is well known to rise steeply from the local universe to $z\sim 1$. \citet{burgarella13} report that the local ratio $LD_{\rm IR}/LD_{\rm UV}=4.25\pm 0.35$ rises by a factor $\sim$3 to reach a peak of $13.44\pm 4.68$ at $z \sim 1.2$, and then falls again, reaching values similar to local at their last data point at $z\sim 3.6$, in accordance with recent evidence from rest frame UV data for decreasing obscuration of the galaxy population at high redshifts. 

Since previous mid-IR based determinations of the IR luminosity function and IR luminosity density used a range of recipes to extrapolate to total IR, there is no single correction that can be applied to them. Comparisons in the references cited indicate that the new \herschel\ based LFs are consistent with or slightly lower than typical \spitzer\ based results. Several past studies reported a discrepancy between
cosmic star formation history and cosmic mass assembly: The cosmic stellar mass densities inferred by integrating the star formation rate densities were a factor $\sim$2 larger than those from stellar mass measurements  \citep[e.g.][]{hopkinsbeacom06,perezgonzalez08}. \citet{burgarella13} argue that these two measurements are now in good agreement, due to a combination of the somewhat reduced \herschel\ based star formation rate densities, the evidence for a decreased average UV obscuration at $z\gtrsim 1.2$ which lowers UV-based SFRs at high redshift, and proper consideration in the comparison of the stellar mass that is returned to the ISM during stellar evolution \citep[see also][]{behroozi13}. The total stellar mass inferred from IR/UV star-formation rates clearly depends on the assumed stellar IMF on the lower main sequence. The improved agreement between inferred and measured total stellar mass thus reduces the need to invoke bottom-light IMFs at high z \citep[e.g.,][]{dave08}.

\subsection{Clustering of infrared galaxies and the role of environment}
\label{sect:clust}

\begin{figure}
\center
\includegraphics[width=\textwidth]{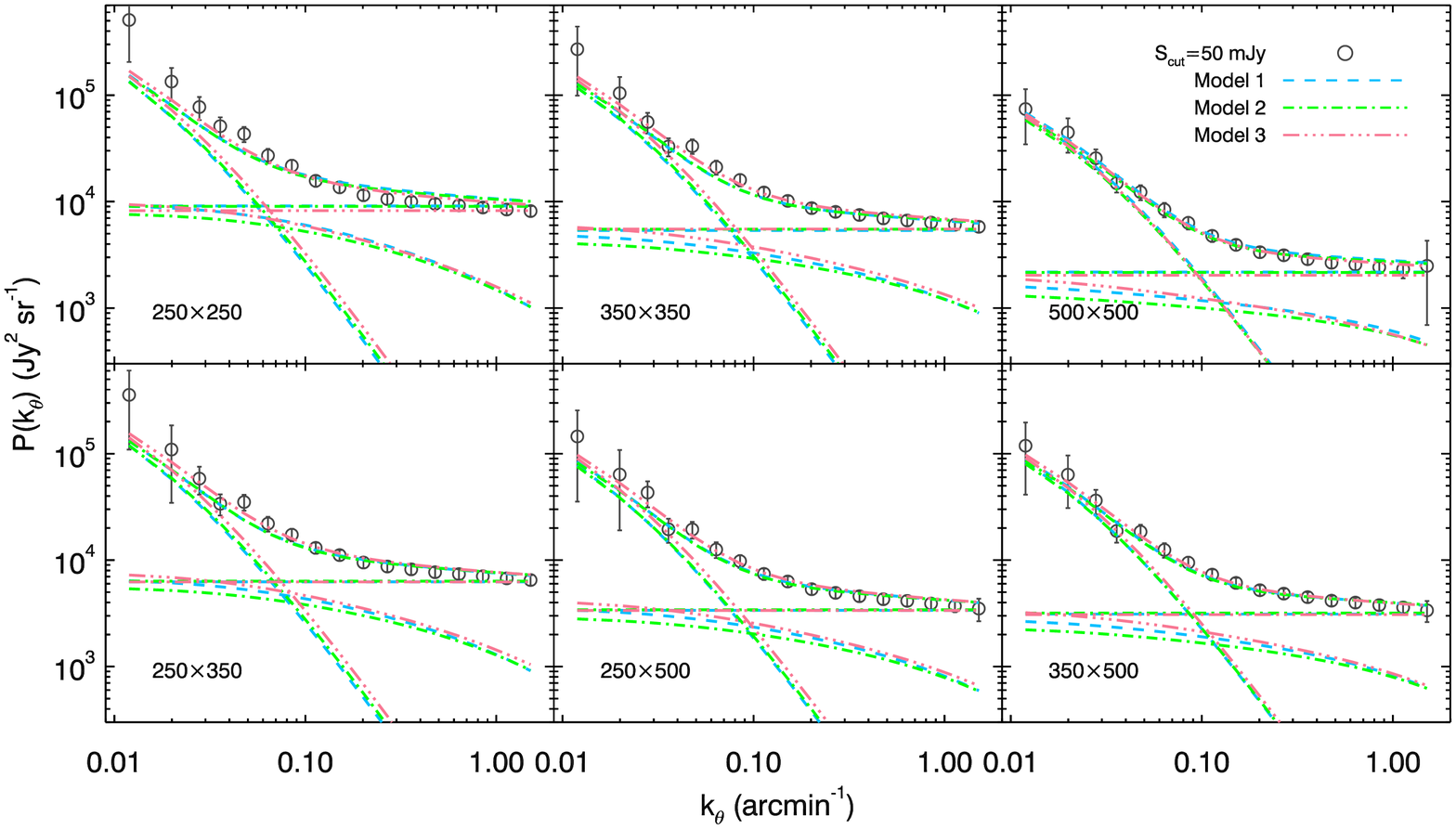}
\caption{SPIRE auto- and cross frequency power spectra, with sources at fluxes greater than $S_{\rm cut}=50$~mJy masked and cirrus subtracted. Three halo models \citep[see][]{viero13a} are fit to these power spectra as well as to the number counts of \citet{glenn10}. Spectra are fit with three terms: Poisson (horizontal lines), 2-halo (steep lines and dominant at low $K_\Theta$) and 1-halo (less steep). Reproduced from Figure~11 of \citet{viero13a}.} 
\label{fig:fluctuations}
\end{figure}

Large survey areas and sample sizes, and improved systematics of current far-infrared surveys provide new opportunities to explore the clustering properties of dusty star forming galaxies and to investigate the impact of environment on their properties.  

Observed and deprojected two-point correlation functions for individually detected
PACS sources are presented in \citet{magliocchetti11,magliocchetti13}, with comoving correlation lengths $r_0$ increasing towards higher redshift and fainter sources. Sources selected at rest frame 60~$\mu$m with similar $SFR\gtrsim 100$~\msunyr\ locally reside in low density environments but at $z\sim 2$ are found in massive $M_{\rm min}\sim 10^{13.5}$~\msun\ halos. For SPIRE selected $z<0.3$ samples, \citet{guo11} and \citet{vankampen12} report $r_0\sim$5\,Mpc similar to $z<1$ PACS sources. Exploiting the gravitational magnification of background Lyman break galaxies, \citet{hildebrandt13} suggest that luminous $S_{250}>15$~mJy SPIRE sources reside in massive log$_{10}(M)=13.2$ halos, in broad agreement with the $z>1$ PACS sources and $z=1-3$ submillimeter galaxies \citep{hickox12}. 

Given significant confusion limits but large area high quality datasets, much of the related analysis using SPIRE as well as \planck\ has focussed on power spectrum analyses of the CIB fluctuations \citep[e.g.,][]{amblard11,planck11,thacker13,planck13b}, with the currently most complete SPIRE analysis presented in \citet{viero13a}. Power spectra and cross-frequency power spectra from SPIRE are available over scales $k_\Theta=0.01-2$ arcmin$^{-1}$, clearly showing the 1-halo and 2-halo clustering terms (Figure~\ref{fig:fluctuations}).

A variety of halo and abundance matching models is under development to match these and other far-infrared data \citep[e.g.][]{bethermin13,wang13,viero13a,planck13b}. Results to date indicate that star formation is generally found to be dominated by $M\sim 10^{12}$~\msun\ halos at all redshifts. These halos represent the progenitors of structures of larger and larger present day mass as one goes back in time. Variations still exist in the prescriptions used by the various models to match halo mass and star formation rate, as well as for the adopted or fitted infrared SEDs, and are worth further scrutiny.

It is well known that local luminous infrared galaxies preferentially reside in underdense regions rather than clusters. \herschel\ studies at $z<0.5$  confirm this
\citep{dariush11,burton13} and suggest that the difference is mostly mediated by a change in the ratio of blue star forming to red passive galaxies, rather than by a change in the intrinsic properties of the star forming galaxies. However, \citet{rawle12b} present evidence for a population of unusually warm galaxies in $z\sim 0.3$ massive clusters, suggested to be objects where cold dust has been stripped from the galaxy's periphery. Comparing the total star formation rate per halo mass ($\Sigma (SFR)/M$) in the field, groups and clusters out to $z>1$, \citet{popesso12} find that groups have low $\Sigma (SFR)/M$ locally, but rapidly catch up with with the field $\Sigma (SFR)/M$ as redshift increases. The cluster $\Sigma (SFR)/M$ increases with redshift but stays below the values for field and groups out to z$\sim$1. It is currently less clear to what extent these changes in the relation of SFR as traced by IR emission and environment, as well as the general rise of star forming activity with redshift, express themselves in a change (or not) of the local relation between galaxy (S)SFR and density \citep{popesso11,ziparo14} or in the level of enhancement of IR emission in galaxy pairs compared to isolated galaxies \citep{hwang11,xu12}.  

\citet{rawle12a} give an overview of star formation rates in brightest cluster galaxies. Use of \herschel\ fluxes avoids potential problems with AGN contamination that could affect previous results that are based on mid-infrared observations. The star formation rates are typically low, 1--150~\msunyr\ even in the far-infrared detected subset of the sample, and comparison to H$\alpha$ based SFRs suggests that star formation is little obscured in these galaxies, and likely spatially very extended. Star formation rates are clearly anti-correlated with the cooling times of the X-ray gas for cool-core clusters, but fall short of nominal X-ray based cooling rates. This is in line with mounting evidence for the importance of AGN feedback in such systems. Only in rare cases does the SFR of the brightest galaxy reach an appreciable fraction of the cooling rate that is suggested by X-ray data \citep{mcdonald12}.

\section{Galaxies and infrared spectral energy distributions in the context of the evolving star forming sequence}
\label{sect:sed}

The power of far-infrared emission as star formation indicator simply relies on the total infrared emission, but the detailed shape of the spectral energy distribution of a galaxy encodes the conditions in the interstellar medium in which the emitting dust grains reside. A dust grain will equilibrate to a temperature that increases with the intensity of the local radiation field. Since the grain's re-emission scales with $T_{\rm dust}^{4+\beta}$ and the dust emissivity index is $\beta$=1.5--2, few K variations around dust temperatures of typically 30~K already reflect significant variations in the average radiation field. Dust heating is a local process that is dependent on the spatial distribution of the heating sources and the dust (acting both as heated and obscuring material), but strongly simplified assumptions can give insights for possible scalings. Compact high pressure and high radiation field intensity star formation regions in the nuclear regions of galaxy mergers will have warmer dust. Such a link between size and temperature can however not be extended to the total size of a galaxy if temperature is driven by the conditions in localized star forming regions, of which more or less may be present in a galaxy's disk but with similar conditions in each. Another simple assumption is to consider the dust as a calorimeter that re-emits the total radiation from young stars. The temperature then reflects the ratio of SFR to dust mass, keeping the ratio of SFR and gas mass fixed (plasible in regimes with a near-linear star formation law), lower metallicity galaxies with lower dust-to-gas ratios will then have warmer dust.   

The dust temperature will also depend on the grain's optical properties as driven by chemical composition, size, and structure. Its emission at this temperature again depends on optical properties via the wavelength-dependent emissivity. Small grains are transiently heated by individual photons, to higher than equilibrium temperature. Such transient heating is important for the continuum on the short wavelength (Wien) side of the far-infrared SED peak, and in particular for the mid-infrared `PAH' emission features which reflect the aromatic nature of their carriers. An example of a current model of interstellar dust reflecting these considerations is \citet{draineli07}. Warmer far-infrared SEDs or colors thus reflect a more intense radiation field, via a long chain involving the properties of the mixed grain population. Additional complication arises because the dust emission may not be optically thin as implicitly assumed above and valid for normal galaxies, but optically thick into the far-infrared. Comparing infrared luminosities, SED temperatures, and sizes to the Stefan-Boltzmann law for blackbodies suggests this is an issue for the most compact local ULIRGs. In contrast, attempts to infer the wavelength of optical thickness from fits to high redshift SEDs should be seen with some skepticism, given that photometric coverage and SNR are typically limited, and the assumptions on the intrinsic SED are often simplified.

Caution is warranted when comparing dust temperatures from different references, since a range of conventions is in use in assigning a single temperature to a complex SED. Most often, dust temperatures are derived by fitting the photometric data with a variant of the modified blackbody function $S_\nu\propto B_\nu(T_{\rm dust})(1-e^{-\tau_\nu})$  with a wavelength dependent optical depth $\tau_\nu=\tau(\nu_0)(\nu/\nu_0)^\beta$. For a given set of fluxes, the derived dust temperature \tdust\ will depend on the adopted dust emissivity index $\beta$ (typically assumed to be 1.5--2), and on adoption of an optically thin model versus a model that is optically thick below some frequency $\nu_0$ (which may be fitted to the data or adopted as a fixed value). Finally, the rest wavelength range included will influence \tdust\ since galaxy SEDs deviate from the single modified blackbody especially on the Wien side of the peak, due to hotter dust and/or transiently heated dust grains. Differences in nominal \tdust\ of order 10\,K can emerge from different conventions. This is significant for typical $T_{\rm dust}=30$~K, and relevant given the good quality and rest frame coverage of current data. Further modifications in use include (1) fits with modified blackbodies at several discrete temperatures or with a power law distribution of temperatures, (2) fits with modified blackbodies joined with a power law on the Wien side, and (3) use of peak wavelengths or frequencies in the fitted SED, converted to a `temperature' via Wien's displacement law.

In the local universe, an extensively studied relation exists between the infrared luminosity of a galaxy and its far-infrared color or dust temperature -- local IR luminous galaxies with their high star formation rates arising in often small regions show warmer dust \citep[e.g.][and references therein]{chapin09,hwang10}. \herschel\ data permit to test whether this \lir\ -- \tdust\ relation evolves with redshift, and how the findings relate to the current picture of high z galaxies in the context of the evolving star forming sequence.

\subsection{Evolution of infrared spectral energy distributions with redshift} 

\begin{figure}
\center
\includegraphics[width=10.cm]{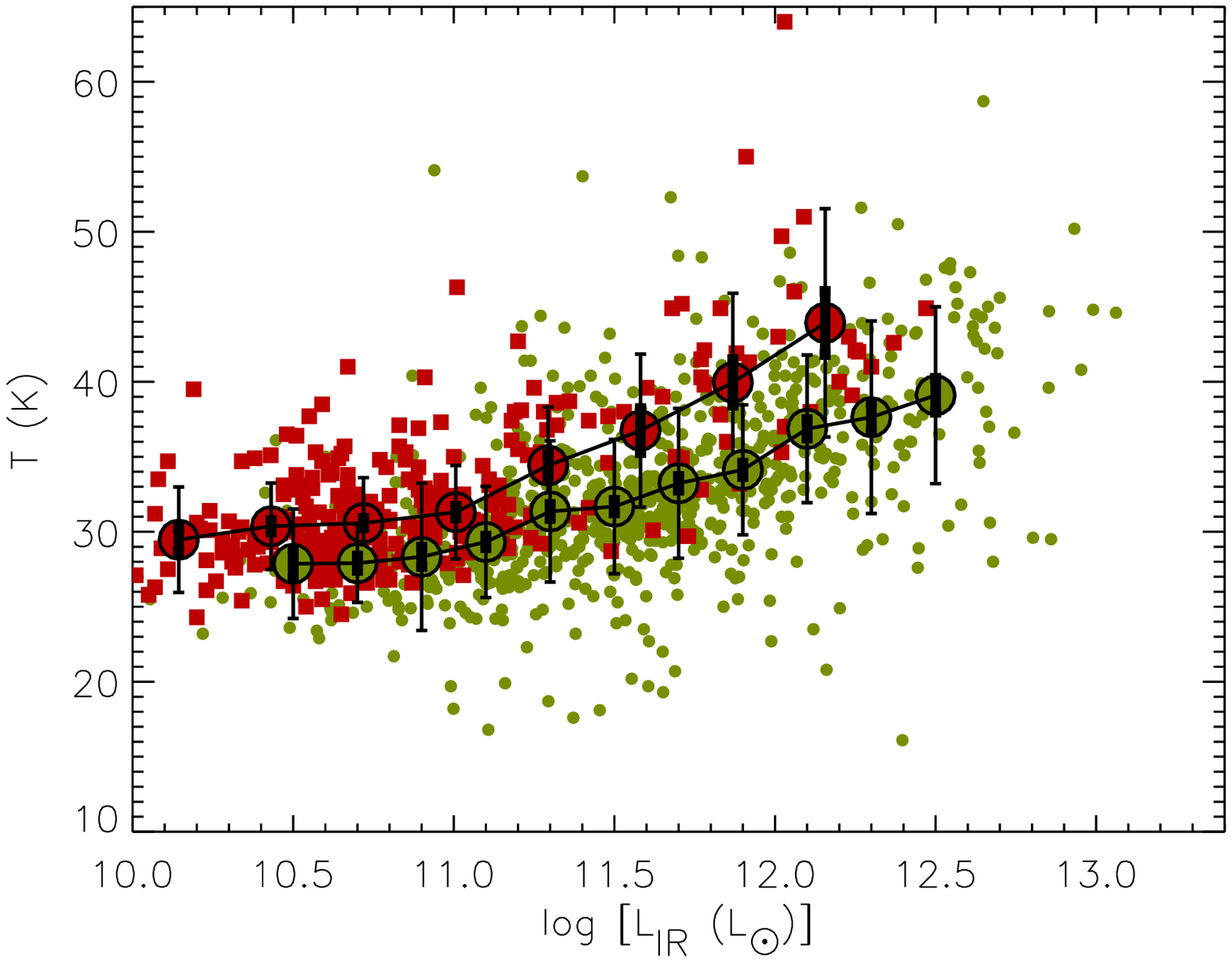}
\caption{\lir\ -- \tdust\ relation for local z$<$0.1 galaxies (red squares) and for a \herschel\ sample (green circles), representing redshifts from z$\sim$0.2 to z$\sim$1.5 as IR luminosity increases. Reproduced from Fig.~18 of \citet{symeonidis13}. 
}
\label{fig:tdlir}
\end{figure}

As discussed in the context of the use of the mid infrared emission as star formation indicator (Section~\ref{sect:recalibsfr}), there is an evolution with redshift in the 
ratio of rest frame 8~$\mu$m PAH emission and total infrared emission. At given IR luminosity, high-z galaxies have relatively stronger PAH, than local ULIRG
mergers. This indicates that they have less intense local radiation fields 
\citep[e.g.][]{tielens08}.

The other main line of studies addresses the dust temperature which is characterizing the rest far-infrared peak. Care is needed due to the common presence of selection effects. In particular SMGs from ground based 850$\mu$m surveys were long expected to be biased towards cold objects. This is confirmed by \herschel\
(Section~\ref{sect:smg}) - there are warmer objects at the same z and $L_{IR}$ that escape the SMG selection. Selection effects certainly also contribute to lower dust temperature and larger dust mass when comparing $z<0.5$ 250~$\mu$m-selected galaxies and local IRAS selected samples \citep{smith12}.

\herschel\ selection near the SED peak is less biased. In a first comparison of $z>0.5$ \herschel\ detections with local galaxies which have SED coverage out to 160\,$\mu$m from \akari\/, \citet{hwang10} conclude that the median $T_{dust}$ of $L_{IR}>5\times10^{10}$~\lsun\ \herschel\ sources is 2--5~K colder than that of local galaxies with similar luminosity, and that the dispersion in dust temperature is larger in the \herschel\ sample. Such a shift to colder SEDs at higher redshift but equivalent IR luminosity is also reported by \citet{rex10}. \citet{symeonidis13} address this issue using a very large SPIRE and PACS dataset. Based on comparison to a wide grid of model SEDs, they restrict the luminosity and redshift range studied to carefully ensure absence of selection effects on the dust temperature. At the z$\gtrsim$1 $L_{IR}\gtrsim 10^{12}$~\lsun\ upper end of the range studied, SEDs are 5-10\,K colder than for local sources of equivalent luminosity (Figure~\ref{fig:tdlir}). High redshift galaxies have lower average interstellar radiation field intensities than their same luminosity local analogs, reflecting more extended star formation.

Placing the infrared SED in the context of the basic parameters stellar mass, star formation rate, and redshift, \citet{magnelli14} derive average dust temperatures from stacked SEDs in bins gridding the SFR -- \mstellar\ plane for six redshift slices out to $z\sim 2$. They derive scaling relations for the dust temperature as a function of these parameters. At all redshifts, \tdust\ smoothly increases with IR luminosity, with specific star formation rate, and with specific star formation rate offset from the star forming main sequence. The classical relation of dust temperature with IR luminosity is less tight than the other two, however, as also evident in the sloped iso-temperature lines in the SFR -- \mstellar\ planes for the various redshift slices (example in bottom right of Fig.~\ref{fig:ms}). Dust temperature is better expressed in relation to the evolving main sequence than in relation to absolute IR luminosity. The scalings derived by \citet{magnelli14} for massive $z\lesssim 2$ star forming galaxies as a function of redshift and SSFR, and as a function of redshift and SSFR offset from the main sequence are: 

\[T_{\rm dust}=98\times (1+z)^{-0.065}+6.9\times {\rm log(SSFR)}\]
\[T_{\rm dust}=26.5\times (1+z)^{0.18}+6.5\times ({\rm log(SSFR) - log(SSFR_{MS})})\]

with similar quality of the two scalings. For normal star forming galaxies on the star forming main sequence, both \citet{magdis12a} and \citet{magnelli14} report a slow increase of dust temperature (radiation field intensity) with redshift. In a simple scenario where dust temperature is directly linked to the ratio of SFR to dust mass as an approximation to radiation field intensity, this may be linked to a combination of a decrease in molecular gas depletion times $M_{Mol}/SFR\propto (1+z)^{-1}$ \citep{tacconi13} and a decrease in metallicity, for main sequence galaxies at increasing redshift. Such trends should then continue to regimes that are not covered by Herschel blank fields surveys. Indeed, for an optically selected  sample of gravitationally lensed $z=1.4-3.1$ $M_*\sim 10^{10}$~\msun\ near main-sequence galaxies that would be too faint for an SED study in blank fields, \citet{saintonge13} report rather warm dust temperatures of order 50~K, plausibly linked to their short gas depletion times $M_{Mol}/SFR$ and their low metallicity.

\citet{magdis11a} explore whether deep \herschel\ surveys uncover a qualitatively new population of galaxies, that is missing in even very deep mid-infrared \spitzer\ data. Indeed there is a small $\sim$2\% fraction of GOODS-S field PACS sources that are undetected even in the deepest 24~$\mu$m data. According to spectroscopic or optical/NIR photometric redshifts, these are found to cluster around redshifts z$\sim$0.4 and z$\sim$1.3 where the 24~$\mu$m band can drop out due to strong silicate absorption. These sources are similar to other compact and obscured starbursts among the sources that are detected at both mid- and far-infrared wavelengths.

\subsubsection{Far-infrared photometric redshifts for large samples}

Identifications and redshifts are relatively easily secured for $z\lesssim 2.5$ \herschel\ galaxies in the best studied fields with deep multiwavelength imaging in particular from \spitzer\/, and with rich sets of spectroscopic and photometric redshifts. Spectroscopic completeness of $\sim$70\% and essentially complete optical/NIR photometric redshifts are in hand for PACS sources in the best fields like GOODS-S. For higher redshifts and in the largest fields with less complete multiwavelength data, it is of interest to directly estimate redshifts for large samples, without a need for identifying counterparts in deep optical/NIR imaging. This is analogous to earlier radio--submillimeter--infrared attempts for submillimeter galaxies \citep[e.g.][]{carilli00,aretxaga07,daddi09}.   
The observed SED peak wavelength can be well measured for large numbers of SPIRE/PACS detected sources, but conversion into a redshift estimate is subject to the well known \tdust --$z$ degeneracy. Given the scatter and evolution with redshift of dust temperature even at given IR luminosity, this is a significant limitation. Specifically, in $z\lesssim 1$ flux limited samples selection effects can conspire to leave very little evolution in the median observed peak wavelength, as shown on the basis of a spectroscopic redshift survey of SPIRE sources by \citet{casey12b}. 

A variety of approaches to \herschel\ photometric redshifts based on rest frame far-infrared data has been reported \citep[e.g.][]{amblard10,dannerbauer10,lapi11}, with an average accuracy of $(1+z_{\rm phot})/(1+z_{\rm spec})\sim 10\%$ in the best reported case \citep{chakrabarti13}. Calibration datasets are often biased towards sources with previous groundbased (sub)mm detections and redshifts, in some cases the method encodes the \lir --\tdust\ relation of SMGs which is biased towards low \tdust\ for  low infrared luminosities \citep{roseboom12a}. Further validation is clearly needed. While \herschel\ SEDs can provide first redshift estimates at $z\gtrsim 1$ or help resolve significant ambiguities in photometric redshifts that are based on other wavelengths, their use for any analysis that needs high quality photometric redshifts should be seen with caution.  

An independent route towards a redshift distribution of \herschel\ sources has been explored by \citet{mitchellwynne12}. Spatially cross-correlating galaxy samples selected at 250 to 500~\mum\ with larger samples selected at optical and near-infrared wavelengths and with known redshift distributions, they infer redshift distributions for the \herschel\ sources with mean $<z>=1.8...1.9$ and broadly consistent with other evidence. Uncertainties are currently large but may improve with inclusion of a larger fraction of the existing \herschel\ data.

\subsection{Infrared galaxies in relation to the star forming main sequence}
\label{sect:ms}

\begin{figure}
\center
\includegraphics[width=13.cm]{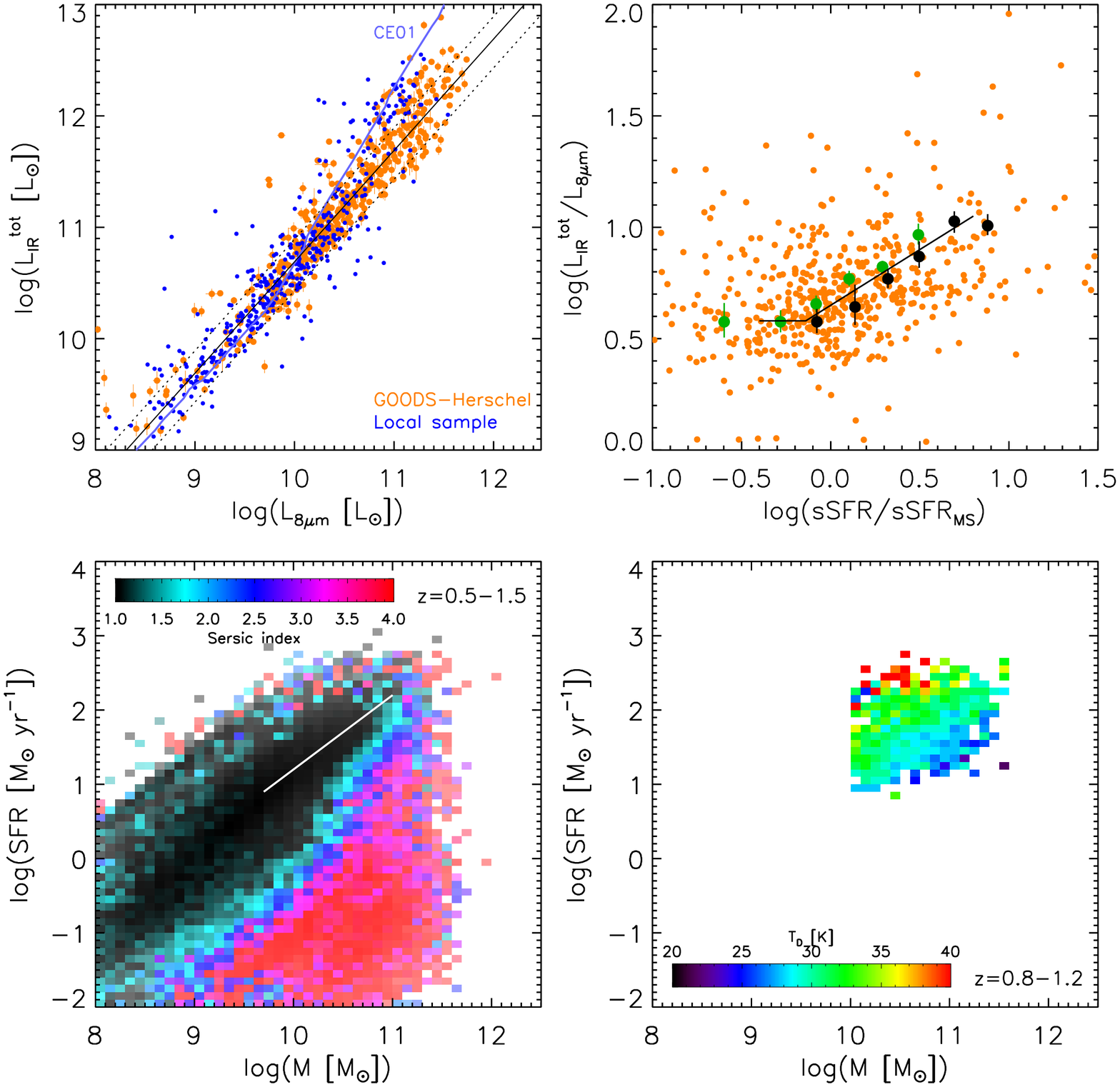}
\caption{Changes in infrared SEDs of galaxies and their relation to the evolving star forming sequence. 
{\bf Top left}: Total infrared vs. 8~$\mu$m luminosity for a local sample (blue dots) and  a $z\lesssim 2.5$ \herschel\ sample (orange dots, diagonal lines visualize median ratio and dispersion for this sample). 
The blue line shows the locus traced by the \citet{chary01} local SED library. Figure adapted from \citet{elbaz11}.
{\bf Top right}: Relation between the offset in specific SFR from the evolving main sequence (`starburstiness') and the ratio of total infrared and 8~$\mu$m luminosity. 
The high-z sample of \citet{elbaz11} is shown in orange. Average values from \citet{nordon12} are plotted for z$\sim$1 (green) and z$\sim$2 (black), along with their proposed redshift-independent relation (black, their Equation 3).
{\bf Bottom left}: Median Sersic indices for galaxies in the $z\sim 1$ star formation rate -- stellar mass plane. Low values indicating preference for disk morphologies are found near the main sequence which is qualitatively located by the white line (adapted from \citet{wuyts11b}).
{\bf Bottom right}: Mean temperature of dust in the $z\sim 1$ SFR -- stellar mass plane (adapted from \citet{magnelli14}).
}
\label{fig:ms}
\end{figure}

\begin{figure}
\center
\includegraphics[width=13.cm]{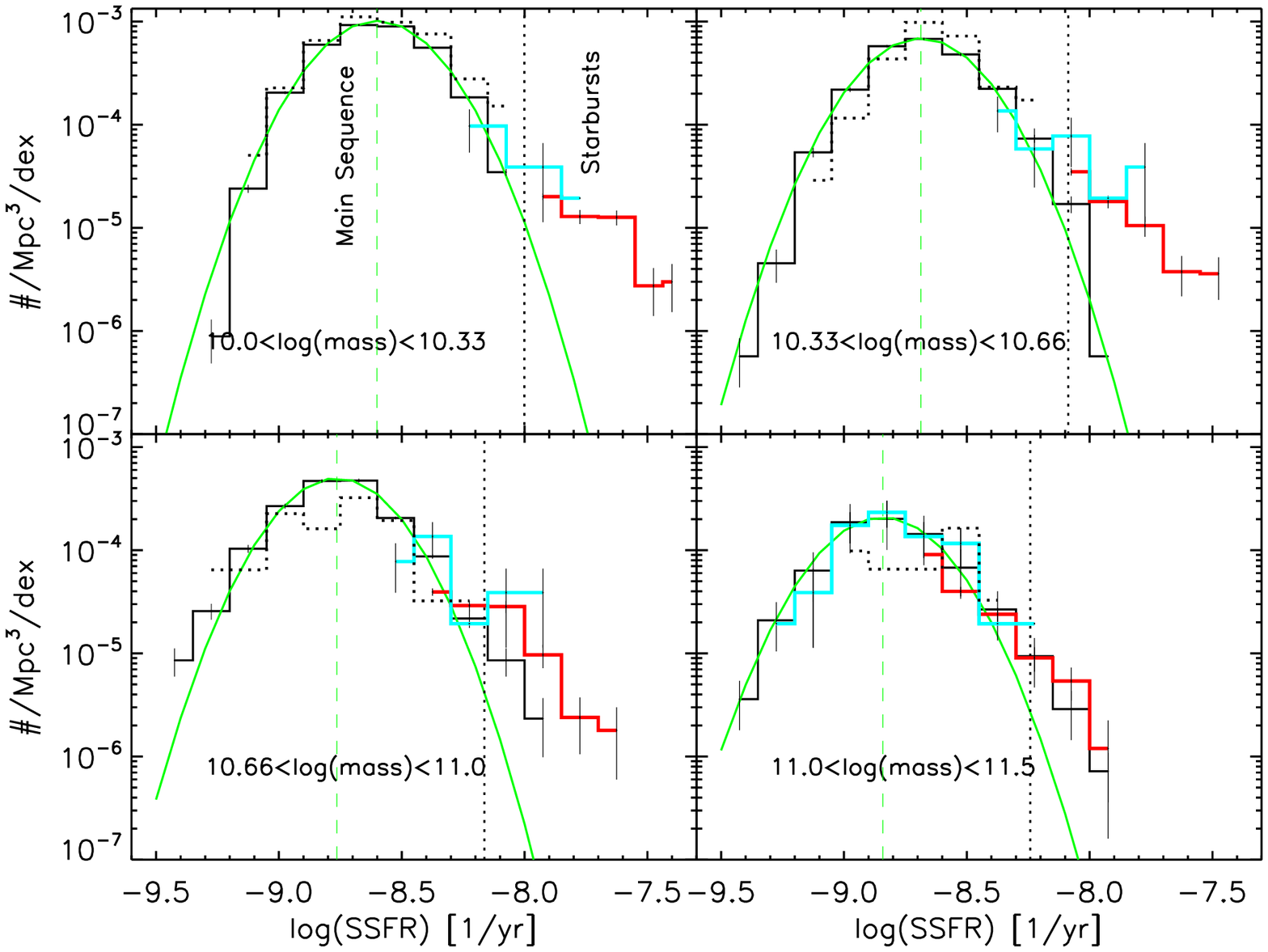}
\caption{Distribution of specific star formation rates for several mass bins in a sample of star forming galaxies at $1.5<z<2.5$.
Star formation rates are from \herschel\ for high SFR sources in COSMOS (red) and GOODS-S (cyan), and from dust-corrected UV emission for BzK galaxies in GOODS-S (dotted black) and COSMOS (solid black). The green curves show the Gaussian fits to the black solid histograms, with the main sequence location at the vertex indicated by the green dashed line. The black dashed lines are 0.6~dex above the main sequence and show the adopted separation between main sequence and starbursts that are off the sequence. Reproduced from Figure~2 of \citet{rodighiero11}.
}
\label{fig:r11}
\end{figure}

Over the last years, star forming galaxies out to at least $z \sim 2$ have been shown to follow a relation between star formation rate and stellar mass, dubbed the `star forming sequence' or `main sequence of star formation' \citep[e.g.][]{brinchmann04,noeske07,elbaz07,daddi07}. Starburst galaxies rise to higher star formation rates above this sequence, and a population of passive galaxies is found at low star formation rates, in analogy to the divison of galaxies into red sequence and blue cloud in color-magnitude diagrams. The star forming sequence is rather tight with a dispersion of $\sim$0.3\,dex, and star formation rate increases with stellar mass with a power law $SFR\propto M_*^p$ with slope $p\sim 0.6-1$. Most importantly, star formation rates of galaxies on this sequence increase with redshift with $(S)SFR\propto (1+z)^{2.5-3}$ to z$\sim$2, with a possible flattening at higher redshift \citep[e.g.][but note ongoing discussion]{gonzalez10}. The existence and tightness of the sequence argue for rather steady star formation histories. This has contributed to the ongoing reassessment of the roles in the driving of star formation that are ascribed to more steady processes such as gas inflow from the cosmic web and minor mergers, versus dramatic major mergers. Here, a close link exists to the finding that many $z\sim 1-2$ massive star forming galaxies are (clumpy) disk galaxies \citep[e.g.][]{foerster09}, with large molecular gas fractions \citep[e.g.,][]{daddi10,tacconi10,tacconi13}. 

The SFR limits of large area \herschel\ surveys reach the main sequence only at its $\gtrsim 10^{11}$\msun\ tip and create samples dominated by starbursts above the sequence. While such surveys provide important statistics for luminous objects, attempts to identify the main sequence from such data alone fail \citep[e.g.][]{lee13}. Studying main-sequence objects at lower masses requires the deepest \herschel\ data e.g. from the GOODS fields, or combination with other star formation indicators, or \herschel\ stacking of star formation samples selected at other wavelengths.

Most star formation happens on the star forming sequence. Combining \herschel\ SFRs for strongly star forming galaxies with optical/NIR data for BzK galaxies with lower star formation rates, \citet{rodighiero11} estimate that only 10\% of the $z\sim 2$ SFR density  is due to objects with SSFRs more than a factor 4 above the main sequence (Figure~\ref{fig:r11}). In numbers, only 2\% of $z\sim 2$ star forming galaxies are found above this threshold. With different parametrisation (two overlapping Gaussian components rather than a threshold separating main sequence and starbursts) the outlying starbursts may contribute $\sim$20\% of the star formation density \citep{sargent12}. Such a predominant role for the star forming sequence is similar to what is suggested by observations in the local universe \citep[e.g.][]{brinchmann04}.

\citet{elbaz11} find that at given IR luminosity, the ratio $\nu L_\nu (8\mu m)/L_{\rm IR}$ increases with redshift (i.e. its inverse, dubbed IR8 by 
\citet{elbaz11}, decreases), causing the mid-infrared excess that affected previous mid-IR based star formation rate estimates (Section~\ref{sect:recalibsfr}). \citet{elbaz11} systematically study this ratio for both a $0<z<2.5$ \herschel\ sample and a local reference sample of star forming galaxies (Figure~\ref{fig:ms} top left). In both samples, the distribution of IR8 is characterised by a prominent Gaussian-like peak centered near IR8$\sim$4, and a tail towards larger values (weaker PAH emission). The peak can be identified with the main sequence of star forming galaxies. In particular for the well-studied local sample, the tail towards larger IR8 is shown to  represent compact starbursts of high IR surface brightness and high specific star formation rate. Changes of IR8 can be related to the main sequence in a way that is consistent over $0<z\lesssim 2.5$. This can be done as two fixed ratios for main sequence objects and for starbursts above the main sequence, separated at IR8=8 and
with IR8=5 and IR8=11 for the derived main sequence and starburst spectral templates  \citep{elbaz11}. Alternatively, $\nu L_\nu (8\mu m)/L_{\rm IR}$ can be expressed as a steady function of the SSFR offset from the main sequence \citep{nordon12}:

\begin{equation}
 \begin{array}{ll}
  \log({\nu}L_{\nu}(8\mu m)/L_{\rm IR}) = \\
  -0.58_{\pm 0.04}                                & : \Delta_{\rm MS} < -0.14\\
  -0.65_{\pm 0.03} - 0.50_{\pm 0.06}\times\Delta_{\rm MS}& : \Delta_{\rm MS} \geq -0.14
 \end{array}
\end{equation} 

where $\Delta_{\rm MS}=log(SSFR)-log(SSFR_{\rm MS})$ is the specific star formation rate offset from the main sequence (Gyr$^{-1}$) (Fig.~\ref{fig:ms} top right. Galaxies with a similar ratio $\nu L_\nu (8\mu m)/L_{IR}$ tend to lie along lines of constant logarithmic SSFR offset from the main sequence in the z$\sim$1 and z$\sim$2 samples of \citet{nordon12}.

The link of these findings to size and morphology is emphasized by the study of \citet{wuyts11b} who use a hierarchy of star formation rates from \herschel\/, \spitzer\/ and UV-optical-NIR SED fitting to place large samples of $z\sim 0.1$ to $z\sim 2.5$ galaxies on the SFR-mass plane and study their HST morphologies. In a median sense, main sequence galaxies show the largest sizes at a given stellar mass, as well as disk-like Sersic indices, while quiescent galaxies as well as starbursts in the upper envelope of the main sequence prefer higher Sersic indices. The starbursts may be in the process of building up central mass concentrations. Interestingly, disk-like morphologies provide a tracing of the main sequence (Fig.~\ref{fig:ms} bottom left) that is independent of the original identification of the star forming main sequence via the correlation of star formation and stellar mass for star forming galaxies. The morphology trends reported by \citet{wuyts11b} refer to the median in a certain SFR, \mstellar , z bin. This leaves room for a mix of disk and merger morphologies within such a bin. Reliable quantification of such fractions is affected by the well known intricacies of accurately distinguishing clumpy single objects from mergers, if using broadband morphological information only.

Concerning the rest frame far-infrared peak, high redshift starbursts above the main sequence show increased dust temperatures \citep{elbaz11,magdis12b,magnelli14}. This is analogous to the difference between warm (U)LIRGs and cold lower luminosity disk galaxies in the local universe. As discussed above, when looking at a broader redshift range dust temperature can be expressed as a function of main sequence offset with less scatter and less redshift dependence compared to expressing temperature as a function of IR luminosity only.

Another key ISM diagnostic is the rest frame far-infrared fine structure lines.
The [CII] 158~\mum\ line, the main interstellar medium cooling line under many conditions, is known since early airborne detections and the ISO mission to show a deficit (decreased [CII]/\lir\ ratio) in local ULIRGs \citep[e.g.,][and references therein]{malhotra01,luhman03}. Decreased photoelectric heating efficiency in intense radiation fields and dust-bounded HII regions likely play a role in shaping this trend. \herschel\ high redshift detections of this line are difficult, but large ground based submm telescopes are stepping in to provide detections, a field that is about to expand vastly with ALMA. At given IR luminosity, high redshift galaxies typically show larger [CII]/\lir\ than local objects \citep[e.g.][]{maiolino09,stacey10,rigopoulou14}. While many of these targets cannot yet be located with respect to the main sequence, due to lack of reliable stellar masses, \citet{graciacarpio11} combine such literature high-z [CII] detections with local \herschel\ measurements. They find that the difference between [CII] deficit at low and high z largely disappears if the deficit is plotted as a function of 
$L_{IR}/M_{Mol}$ rather than \lir\/. This ratio of IR luminosity and molecular gas mass is a proxy to offset from the main sequence. \citet{graciacarpio11} find that sources showing a deficit in [CII] and also in other far-infrared fine structure lines are found at $L_{\rm FIR}/M_{\rm H_2} \gtrsim 80$~\lsun\ \msun$^{-1}$.

To summarize these findings for massive $M_\ast\gtrsim 10^{10}$\,\msun\ star forming galaxies: For a wide set of interstellar medium diagnostics (PAH emission, far-infrared dust temperature, fine structure line emission) as well as morphology, which are all known to locally follow well characterised trends with infrared luminosity from disk galaxies to merger ULIRGs, the main sequence paradigm seems to provide a better and more consistent reference to express properties in a redshift independent way. Star formation in main sequence galaxies appears to occur in a similar mode at all redshifts, despite a variation in (S)SFR of  more than a factor 20 for main sequence galaxies over the redshift range studied. Compression of the gas content for example in major mergers changes the ISM conditions and enhances the specific star formation rates of starbursts above the main sequence.

\section{Far-infrared emission of high redshift galaxy populations}

\subsection{Gravitational lenses as detailed probes of the interstellar medium at high redshift}
\label{sect:lensed}

\begin{figure}
\center
\includegraphics[width=13.cm]{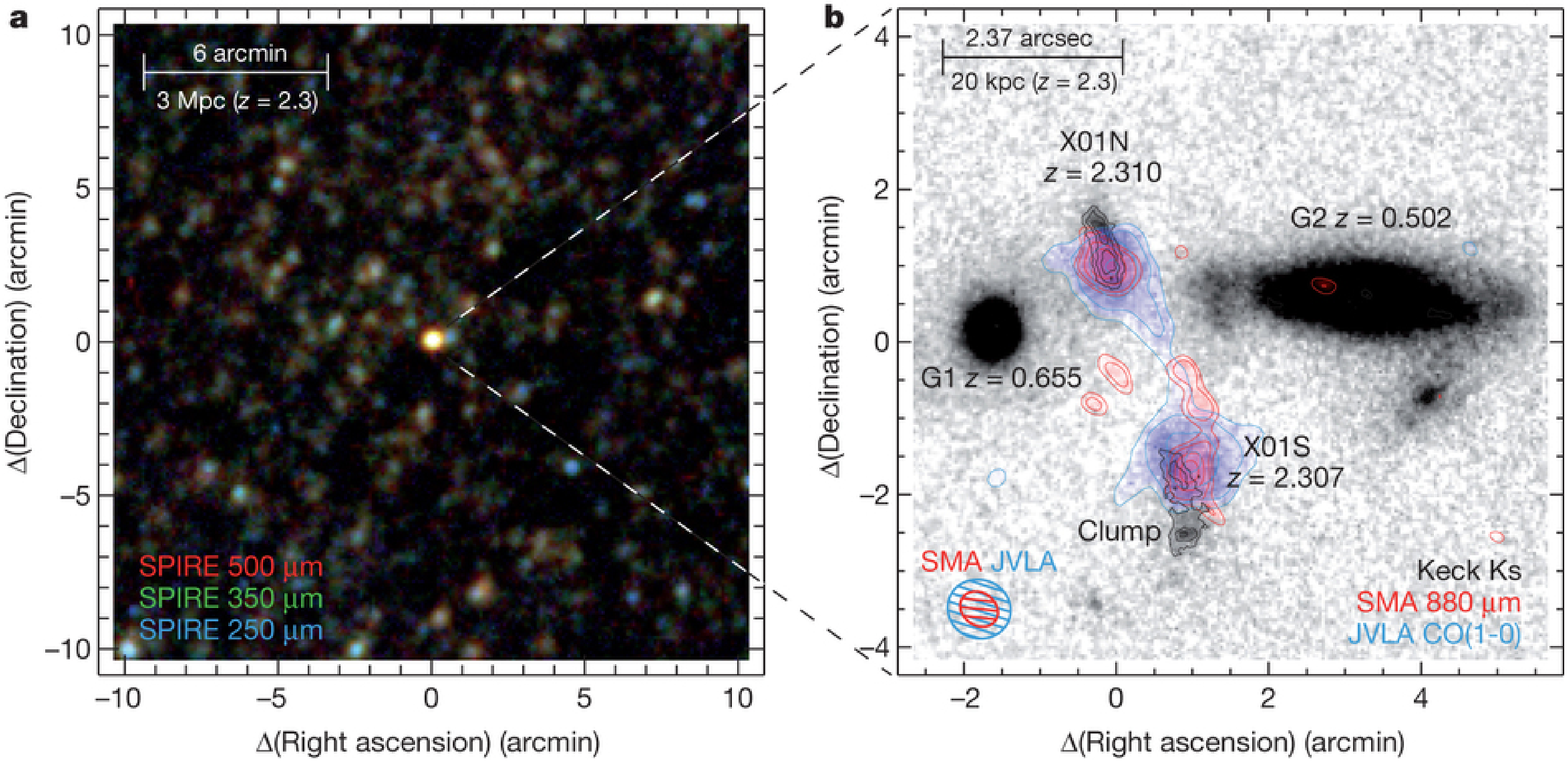}
\caption{Followup of bright \herschel\ 500~$\mu$m or SPT millimeter sources uncovers a large fraction of lensed high redshift galaxies. Properties range from only weakly magnified, intrinsically extremely luminous targets as in the example of the weakly amplified galaxy pair HXMM01 (Figure reproduced from \citet{fu13}), to strongly magnified targets. The left panel shows a SPIRE three color image of HXMM01, the right panel high resolution follow up in near-infrared stellar continuum, submm dust continuum and JVLA CO emission.}
\label{fig:hxmm01}
\end{figure}

Gravitational lensing magnification by individual foreground galaxies or by clusters of galaxies permits detailed studies that are impossible for equivalent sources in blank fields, due to sensitivity limitations. Starting with the detection of IRAS~F10214+4724 in a redshift survey of the IRAS catalog \citep{roro91}, a handful of lensed $z>2$ sources have become accessible for in-depth infrared to millimeter studies. The first few such objects were detected through a variety of techniques that often selected systems hosting powerful AGN. Early on, a number of studies realized that large area surveys at submillimeter and millimeter wavelengths will provide an efficient and less biased method for detecting lensed dusty galaxies at high redshift \citep[e.g.][]{blain96,negrello07}. At these wavelengths, the combination of the steep decay of unlensed source counts towards bright fluxes and the approximate constancy of source flux with redshift due to the negative K-correction on the Rayleigh-Jeans side of the far-infrared SED ensures that a significant fraction of bright sources in a large area survey will be lensed high redshift galaxies, rather than local objects. This differs from optical and X-ray wavelengths, where lensed objects are relatively faint compared to the multitude of foreground sources, due to positive K-corrections.
   
\herschel\/-SPIRE surveys and ground based South Pole Telescope (SPT) mm surveys have now covered about 1000 and 2500 square degrees respectively, sufficient for application of these methods. Flux cuts at $S_{500\mu m}=100$~mJy or $S_{1.4mm}=20$~mJy produce samples that contain candidate lensed dusty galaxies at the tens of percent level. Separating these from radio galaxies, low redshift galaxies, and galactic dust clouds at the same flux level is relatively straightforward, based on (sub)mm colors as well as readily available radio and optical data \citep{vieira10,negrello10}, see also \citet{marsden13} for a search from Atacama Cosmology Telescope data. Hundreds of lensed galaxies are expected to be present at these levels in  the existing surveys, given surface densities of candidate lensed galaxies of 0.14 to 0.26$\pm$0.04 deg$^{-2}$ at $S_{500\mu m}>100$~mJy \citep{wardlow13a,bussmann13} and $\sim$0.1 deg$^{-2}$ at $S_{1.4mm}>20$~mJy \citep{vieira10}. Refined selection methods including color information may push the number of candidates to more than a thousand \citep{gonzaleznuevo12}. Related searches for somewhat fainter sources with extremely red SPIRE colors \citep{dowell14} have the potential of uncovering highest redshift dusty star forming galaxies \citep[e.g.][]{riechers13}, likely with a smaller fraction of lenses. Firming up the census of z$>$4 sources from these searches may tighten the contraints that very high star formation rate high-z sources place on models of galaxy evolution, beyond those known from the z$<4$ SMG and \herschel\ populations. 

Most redshifts of these lensed dusty star forming galaxies have been obtained from direct CO detections. CO lines (or alternatively rest frame mid-infrared features observable with space telecopes) are quite closely linked to the rest frame far-infrared emission that defines these targets. Their direct detection largely avoids the effort, mis-identification risks and biases that are inherent to the traditional way of obtaining first a radio- or mm-interferometric accurate position, then identification in the optical or near-infrared, and finally an optical spectroscopic redshift of the counterpart. The high flux of these lensed objects and the recent availability of wideband spectrometers at both single dish telescopes and interferometers have contributed to a rather rapid buildup of redshift information \citep[e.g.][]{negrello10,frayer11,cox11,scott11,combes12,harris12,lupu12,weiss13,bussmann13}. Spectroscopic redshift completeness well exceeds 50\% for the best studied lensed samples.

\herschel\/- and SPT-detected systems include some of the highest redshift $z \sim 6$ dusty star forming objects that are currently known and do not host a QSO \citep{riechers13,weiss13}, and have already been followed up in a variety of other tracers of the warm and dense interstellar medium \citep[e.g.,][]{omont11,cox11,omont13,rawle13,bothwell13}. Corrected (where known) for the lensing magnification they typically show very high star formation rates ranging from several hundred to few thousand solar masses per year \citep[e.g.][]{negrello13}, including extreme cases that reach $\gtrsim$2000~\msunyr\ and are single galaxies or physically associated pairs according to interferometric followup \citep[e.g.][]{fu12,riechers13,fu13,ivison13}. 
These last cases have passed the detection thresholds simply by their huge luminosities, with only weak amplification suggested by the detailed followup (e.g. Fig.~\ref{fig:hxmm01}).
In their intrinsically large luminosities ($L_{IR}\sim 1.5\times 10^{13}$\,\lsun\/), compact sizes ($R_{half,880\mu m}\sim 1.5$\,kpc) and warm dust temperatures (\tdust$\sim 39$\,K), the lensed sources studied by \citet{bussmann13} resemble high luminosity blank field SMGs, with the effect of lensing biases still under study.

Placing these sources in the context of other galaxy evolution studies is slowed down by the need to construct reliable lensing models on the basis of $<1\arcsec$ resolution data either from (sub)mm interferometry, or near-infrared imaging from HST or adaptive optics, even if methods for crude magnification estimates on the basis of line widths have also been proposed \citep{harris12}. Of the $\sim$30 published lensing magnifications, many are based on $\sim 0.5\arcsec$ resolution data, coarser than desirable given the likely presence of differential lensing effects especially for objects that are strongly amplified by galaxy lenses or near the caustics of cluster lenses. In a statistical sense, lenses from flux limited wide area submm surveys prefer intrinsically compact sources \citep{hezaveh12} with comparison of the expected magnification distributions to observations going on \citep{bussmann13}. This size bias will also affect the ISM conditions. The difficulty of separating lens and lensed object at optical/near-infrared wavelengths severely limits the number of lensed dusty star forming galaxies from wide area searches for which key host parameters such as the stellar mass are currently known. Cases with published stellar mass \citep[][see also \citet{swinbank10}]{fu12,fu13,negrello13} mostly lie above the main sequence, resembling bright blank field SMGs also in that respect.

\subsection{UV selected galaxies}
\label{sect:lbglae}
Over the last two decades, galaxies selected in the rest frame UV via either 
broad band photometry and the Lyman break dropout technique (Lyman break galaxies, LBGs) or by detection as Lyman~$\alpha$ emitters (LAEs) in narrowband searches have played a pivotal role for our understanding of $z\gtrsim 3$ galaxy evolution. An important uncertainty concerns the obscured fraction of their star formation rate. One typical way of deducing obscuration (often expressed as the infrared excess $IRX=L_{\rm IR}/L_{\rm UV}$) is from the slope $\beta$ of the rest frame UV continuum ($f_\lambda\propto\lambda^\beta$), applying locally calibrated relations \citep[e.g.][]{meurer99}. Alternatively, SED fitting techniques are applied to the same plus longer wavelength photometric information. Direct detection of rest frame far-infrared or submillimeter emission from these systems turned out to be difficult \citep{chapman00}. Concerning typical members of the population, such detections have been restricted mostly to a few lensed systems \citep[e.g.][]{baker01,siana09}.

\subsubsection{Lyman break galaxies}

The direct detection of typical blank field LBGs remains out of reach of \herschel\ data. In large samples, a rather small fraction is individually detected but biased towards LBGs that are more massive and have redder rest frame UV colors, for both GALEX-selected $0.7<z<2.0$ LBGs \citep{burgarella11,oteo13b} and for $z\sim 3$ LBGs \citep{oteo13a}. These \herschel\/-detected LBGs have large IR luminosities $L_{IR}>10^{11}\ldots 10^{12}$\,\lsun\/. For larger $z\gtrsim 3$ LBG samples (sometimes pre-selected for larger expected SFR), even stacking analyses are just able to sometimes achieve detections in \herschel\ or groundbased (sub)millimeter data \citep{magdis10b,rigopoulou10,lee12,davies13}. The SED information that can be extracted is limited given the modest S/N of the stacks, but in some cases (unless affected by sample contamination) suggests moderately cold dust temperature \citep[$T_{\rm dust}\sim 30-40$~K,][]{lee12,davies13}. Perhaps the best SED constraints for $z\lesssim 3$ LBG-like galaxies come from small numbers of gravitationally lensed systems, where \citet{saintonge13} find warm dust temperature reaching $T_{\rm dust}\sim 50$\,K \citep[see also][]{sklias13}. There remains uncertainty concerning the rest frame far-infrared SEDs of $z>3$ LBGs. This implies limitations for interpreting detections that ALMA will obtain on the long wavelength tail of the SED, in terms of the balance of obscured and unobscured star formation.

\subsubsection{The IRX-$\beta$ relation at high redshift}
 
In the absence of infrared data, obscured star formation has to be quantified from the rest frame UV emission. Discussing and testing at high redshift the derivation via the IRX-$\beta$ relation mentioned above is of relevance also for more complex SED fitting technigues that are making use of the same photometric information and similar underlying concepts. An empirical IRX-$\beta$ relation such as the classical one of \citet{meurer99} for local UV-selected starbursts (and many revisions since) encodes the properties of the unobscured stellar population such as star formation history and metallicity, and an attenuation law. The attenuation law is sensitive not only to the properties of the obscuring dust that determine the classical extinction curve, but also to the geometric configuration of stars and obscuring dust. Inevitably, stellar emission and dust obscuration in a galaxy will be spatially mixed to some extent. In this approach, however, attenuation is formally treated as a foregound screen. 

Several papers have used \herschel\ data to test the IRX-$\beta$ relation outside the local universe. A main finding is the wide spread in the IRX-$\beta$ plane, also depending on the specific sample studied, which makes it impossible to distill a unique and reliable recipe to estimate the obscured star formation. A sample of $z<0.3$ 250~$\mu$m selected galaxies \citep{buat10} is found to spread from roughly the IRX-$\beta$ relation for local starbursts to that of local normal star forming galaxies, with IRX varying by at least an order of magnitude at given $\beta$. A similarly large spread is observed for $1<z<2.5$ PACS detections \citep{buat12,nordon13}, and in a number of other studies \citep{wijesinghe11,oteo12a,oteo13a,oteo13c}.

Analysing \herschel\ stacks of individually undetected $z\sim 2$ and $z\sim 1.5$ UV selected galaxies, \citet{reddy12} and \citet{heinis13} find that the IRX for such stacks increases with $\beta$, similar to the trends for local galaxies. There is, however, a noticeable factor $>2$ difference between these two studies in the absolute IRX level at given $\beta$, reflected in a better match to local starbursts \citep{reddy12} vs. a better match to local normal star forming galaxies \citep{heinis13}. A similar spread may be indicated in the $3.3\lesssim z\lesssim 4.3$ stacking study of \citet{lee12}.

\citet{buat10} report a decrease in IRX for z$\sim$1 galaxies compared to local galaxies of the same infrared luminosity. \citet{nordon13} observe that the offset of $1<z<2.5$ PACS galaxies from a mean IRX-$\beta$ relation correlates strongly with the offset from the star forming main sequence, with larger IRX above the main sequence. This finding likely relates to the offset from a standard IRX-$\beta$ that is found for local dusty ULIRGs \citep{goldader02} as well to the connection that has been found locally between the IRX-$\beta$ relation and the stellar birthrate parameter \citep{kong04}. UV obscuration properties may be closely connected to the position of a galaxy with respect to the evolving main sequence, in analogy to the infrared SEDs (Section~\ref{sect:sed}).

In their detailed study combining short wavelength and \herschel\ data, \citet{buat12} report detection of a 2175\AA\ UV bump in the attenuation curves for 20\% of their sample, as well as a diversity of attenuation law slopes. Both findings might relate to variations between the galaxies in dust geometry as well as dust properties.

Obscuration of star forming galaxies (IRX) increases with stellar mass \citep[e.g.][]{wuyts11b,buat12,heinis14}. Grouping galaxies detected in both IR and UV by UV luminosity yields a trend in the sense that the most FUV luminous galaxies avoid large IRX \citep{burgarella11,buat12,oteo13b}. Selection as well as evolution may play a role here, and no obscuration trend is obvious in stacks of z$\lesssim$2 UV-selected galaxies that are grouped by LUV \citep{reddy12,heinis13}. Some decrease of IRX towards high LUV  may be present for stacks of 
z$\sim$3--4 UV selected galaxies \citep{heinis14}.

In summary, there is a large spread in the IRX-$\beta$ properties of \herschel\ detected galaxies. Combined with the tendency for underpredicting the SFRs of the most star forming galaxies when using rest UV-optical SED fitting \citep{wuyts11a} or
a local IRX-$\beta$ relation \citep[e.g.][]{oteo13a,oteo13c}, this finding supports to use a hierarchy of SFR indicators, where SFR is based on IR (plus unobscured UV where available) for the most heavily star forming systems detected with \herschel\ or in the mid-infrared, and SFR is based on rest frame UV to NIR SED fitting for lower star formation rate systems (Section~\ref{sect:recalibsfr}).   

\subsubsection{Lyman~$\alpha$ emitters}
Detection of dust continuum emission from $z\gtrsim 5$ Ly$\alpha$ emitters is firmly in the regime requiring \alma\/, although an initial upper limit indicates a complex picture \citep{ouchi13}. At lower redshifts, insights can be gained from single dish submillimeter telescopes or \herschel\/. A small subset of $z\sim 2-3.5$ LAEs is detected by \herschel\ \citep[e.g.][]{bongiovanni10,oteo12b}, in another z$\sim$2.8--4 study there are no individual detections \citep{wardlow13b}. Finding a few LAEs to be IR-bright is in line with the earlier realisation of significant Ly$\alpha$ emission from many SMGs \citep{chapman05}. These individual identifications with dusty star forming galaxies form a biased subset of LAEs, though, calling for stacking analyses or study of local analogs. At z$\sim$0.3, \herschel\ detection rates of \galex\/-selected Lyman~$\alpha$ emitters are more favorable \citep{oteo11}, and LAEs are found to be biased to lower obscuration compared to similarly UV-luminous systems that are lacking Lyman~$\alpha$ emission \citep{oteo12a}. In all properties, classical LAEs selected from narrow band Ly$\alpha$ searches should be contrasted with the extremely IR-luminous dust-obscured population that was recently retrieved from a \wise\ selection by \citet{bridge13}, which also exhibits prominent and extended Ly$\alpha$ emission.

\subsection{Submillimeter galaxies from ground-based surveys}
\label{sect:smg}

Groundbased 850~$\mu$m and millimeter observations provided the first detections of luminous $z\sim 2$ infrared galaxies. These `submillimeter galaxies (SMGs)' continue to be of interest, with ever larger samples being obtained with improving detector technology at ground-based single dish telescopes. Reproducing such a significant high redshift population with star formation rates reaching the regime of 1000~\msun\ yr$^{-1}$ has long been known to be a challenge to theoretical models of galaxy evolution \citep[e.g.][]{baugh05,dave10}. Since SMG luminosities traditionally had to be extrapolated from the submillimeter and radio or from the mid-infrared using locally calibrated relations, confirmation of these large luminosities is important.

For the difficult task of identifying short wavelength counterparts to SMGs, \herschel\ photometric surveys can provide better spectral continuity between the detection wavelength and mid- and near-infrared counterparts \citep{dannerbauer10}. But given \herschel\ beamsizes, they cannot replace radio and in particular (sub)mm interferometry for reliable identification and for characterizing the fraction of single dish detections that breaks up into several sources, either physically associated and possibly interacting galaxies, or chance projections \citep[e.g.][]{hayward11,barger12,smolcic12,karim13,hayward13}.

PACS and SPIRE studies sample the SED peaks of bright $z\lesssim 4$ SMGs with good S/N \citep{magnelli10,chapman10,magnelli12a,swinbank13}. These results strongly confirm the dust temperature selection effects that ground-based SMG samples suffer at given IR luminosity and redshift. In particular for $z\sim 1$, \lir$\gtrsim 10^{12}$\lsun\ galaxies, an object has to be colder than representative for bolometrically selected samples, in order to be detectable in a typical groundbased submm survey. These selection effects are less severe for the most luminous $z\sim 2$ SMGs. \herschel\ data also confirmed that a large fraction of previously known `optically faint radio radio galaxies' (OFRGs) with radio continuum indicative of strong star formation but nondetection in submm surveys are indeed strongly star forming objects but with warmer dust temperatures \citep{magnelli10, chapman10}. A further confirmation of the dust temperature biases in the SMG population stems from the comparison to galaxies selected solely via the stellar rest frame 1.6\,$\mu$m bump. These extend to warmer dust temperatures than SMGs, at same redshift and IR luminosity \citep{magdis10b}.

The huge luminosities of SMGs previously extrapolated from other wavelengths are broadly confirmed by \herschel\/. As for other high redshift galaxies (Section~\ref{sect:recalibsfr}) IR luminosities estimated from 24~$\mu$m emission via locally calibrated luminosity-dependent templates are too high and show large scatter 
\citep{magnelli12a}. Luminosity estimates based on radio and submm flux, redshift, and adoption of the local radio--far-infrared correlation do better, but also require downward revision for the SMG sample with spectroscopic redshifts that was studied with \herschel\ by \citet{magnelli12a}. It would however be premature to interpret this as evidence for evolution of the radio--far-infrared correlation, since these particular SMGs will be somewhat biased to high radio fluxes due to the role radio interferometry plays in the traditional SMG identification and redshift determination process. In line with this, the small SMG sample of \citet{barger12} with full submm and radio identification is consistent with the local radio-FIR correlation.

A picture emerges in which traditional SMG samples that are selected over just a small $S_{850}\sim 3-10$~mJy range are in fact rather heterogeneous \citep[e.g.,][]{magnelli12a}. Such samples cover two orders of magnitude in total IR luminosity, range from objects near the star forming main sequence to more than an order of magnitude above it \citep[but note discussion on SMG stellar masses,e.g.][]{michalowski12}, and cover almost a factor of three in dust temperature. In studying the detailed nature of SMGs and their interstellar medium, it is hence key to differentiate these groups rather than globally refer to `SMGs'. A z$\gtrsim$1 $L_{\rm IR}\sim 10^{12}$~\lsun\ sample \citep[as, e.g., studied in {[}OI{]}\,63$\mu$m by][]{coppin12} will almost certainly differ in its properties from ten times more luminous galaxies at $z\sim 2.5$, despite sharing detection in a ground based SMG survey. This caveat will become ever more important with improving single dish submillimeter surveys and with \alma\ detections of significantly fainter objects at similar wavelengths. The heterogeneity of the SMG population is further emphasized by the analysis of \citet{hayward13}. They use a semi-empirical model based on evolving stellar mass functions and scalings for star formation rate and gas fractions, combined with hydrodynamical and radiative transfer simulations for isolated disks and mergers, to model the SMG population. Number counts are fit without a need to invoke a top-heavy IMF. At faint fluxes $S_{850}\sim 1$~mJy, isolated normal star forming disks and pairs that are blended by the large single dish beam dominate somewhat expectedly. But isolated disks and blended pairs still do so at $S_{850}\sim 3$~mJy, and even at the $S_{850}\sim 10$~mJy bright end, only 2/3 of the SMGs are suggested to be luminous mergers. In going from single dish to interferometric studies, number counts should then be reduced by as much as a factor 2 as blended pairs break up, as also suggested by recent observational studies \citep{barger12, smolcic12, karim13}.

\subsection{Dust-obscured galaxies}
\label{sect:dog}

A considerable portion of \spitzer\ observations and follow up has been devoted to a population of $z\sim 2$ dust-obscured galaxies (DOGs), defined by a high ratio of mid-infrared to rest UV (observed R band) flux density $S_{24\mu m}/S_R\gtrsim 1000$ \citep[e.g.][]{dey08}. At the $S_{24\mu m}\gtrsim 1$\,mJy bright end, these samples include a large fraction of obscured AGN, while at fainter levels often obscured star formation is suggested, from the presence of a rest near-IR stellar bump rather than an AGN power law continuum.

Several studies address the far-infrared properties of this population and are in broad agreement with this general picture (see also the related discussion of a bright 24~$\mu$m selected sample by \citet{sajina12}). Herschel detects half of the $S_{24\mu m}>100$\,$\mu$Jy $z\sim 2$ dust obscured galaxies, at average infrared luminosity $2.8\times 10^{12}$\,\lsun\ \citep{calanog13}. Sources with bump rest frame near-IR SED require a star formation dominated overall SED, while sources with power-law near-IR SEDs are better fit with SEDs such as that of Mrk\,231, including a strong AGN component \citep{melbourne12}. For a sample of dust obscured galaxies that emphasises the faint mostly star forming end of the population (median $S_{24\mu m} = 161$~$\mu$Jy), \citet{penner12} analyse the far-infrared to mid-infrared flux ratio. This ratio is found to be similar as in other dusty star forming galaxies at these redshifts, suggesting that the large mid-IR to rest UV ratio is not due to an extra mid-infrared component but due to atypically large obscuration of the star forming regions in the rest frame UV.

The WISE color-selected population with strong Ly$\alpha$ emission studied by \citet{bridge13} appears to form a yet more extreme end of the dust obscured population, with extreme $L_{IR}>10^{13}-10^{14}$\lsun\ luminosities and warm dust temperatures, and likely a dominant AGN.

\section{Star formation in hosts of active galactic nuclei}
\label{sect:agnhosts}

\begin{figure}
\center
\includegraphics[width=\textwidth]{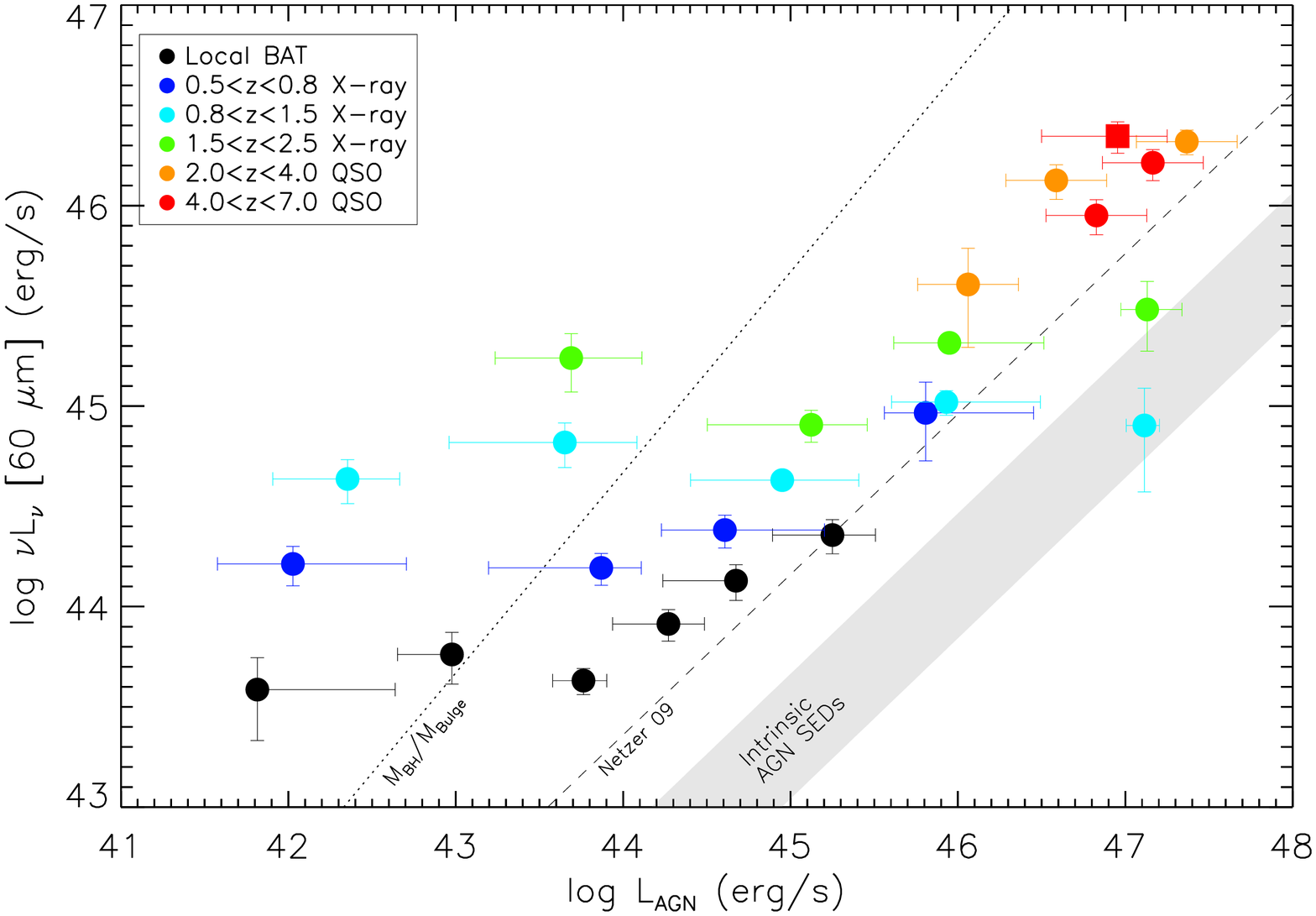}
\caption{Growth of galaxies and their black holes. Average rest frame far infrared emission of AGN hosts expressing the star formation rate is plotted as a function of bolometric AGN luminosity, in different redshift bins. Data for local BAT AGN
and for $z<2.5$ X-ray selected AGN are reproduced from \citet{rosario12}. Data for optically selected high redshift QSOs are from \citet[][and priv. comm.]{serjeant10}
and the z$\sim$4.8 sample of \citet[][square]{mor12b}. Far-infrared luminosities are mean values that include direct detections as well a stacked nondetections. They include all AGN in a bin, and are plotted at the median AGN luminosity with horizontal error bars showing the range including 80\% of the bin's sources. The dotted line indicates the proportionality for a continuous host and black hole growth that would produce the local universe relation between black hole mass and bulge mass \citep[from][assuming black hole accretion efficiency of 0.1]{haering04}. The diagonal dashed line is the correlation for local AGN-dominated sources as proposed by \citet{netzer09}, and the diagonal grey band the approximate 1$\sigma$ range exhibited by empirical pure AGN `intrinsic' SEDs (see \citet{rosario12} for details).} 
\label{fig:growth}
\end{figure}

Ideally, co-evolution of supermassive black holes and their host galaxies should be studied in samples with measurements for all of the four key quantities that capture the status and growth of the supermassive black hole (BH mass and accretion rate) as well as the host galaxy (assembled stellar mass and star formation rate). Concerning host properties, many of the popular star formation indicators (rest frame UV-optical SED fitting, optical emission lines, mid-infrared flux) are significantly affected by AGN driven emission, in particular for type 1 AGN. Since in the rest frame far-infrared the contrast between the typical SED of a star forming galaxy and an AGN SED is the largest, many workers have exploited \herschel\ surveys to study star formation in the hosts of high redshift AGN. One goal of these surveys is to search for correlations between phases of intense host and black hole growth that may occur in the `merger scenario' \citep[e.g.][]{sanders88,dimatteo05} where major galaxy mergers play a strong role in shaping the local scaling relations. Another one is to test to what extent (on the other hand) AGN are hosted by galaxies with already quenched star formation.

Even in the far-infrared part of the SED, the host dominance is not granted. However, to explore the conditions where the host is dominant, theoretical models of AGN heated dust are of somewhat limited value. The freedom to distribute dust outside the sublimation radius in different ways leads to a large variety of warm to cold SEDs, not all of which are actually present around real AGN. Two different empirical approaches have been pursued instead. \citet{netzer07}, \citet{mullaney11} and \citet{mor12a} have decomposed the total AGN+host SEDs of local systems using \spitzer\/-IRS spectra, and spectral templates of star forming galaxies. The resulting `intrinsic AGN SEDs' drop steeply from the rest mid- to far-infrared. \citet{hatziminaoglou10} and \citet{rosario12} have compared infrared colors of high redshift AGN hosts with inactive reference samples, finding the mid-infrared flux often boosted by AGN emission but rest frame far-infrared colors similar in AGN hosts and reference samples. Both approaches suggest that for $z\lesssim 3$ AGN with luminosity similar to typical L$_{2-10keV}\lesssim 10^{45}$ X-ray deep field sources, the rest frame far-infrared emission is on average dominated by the host. This may not be the case for some of the most luminous AGN, or for AGN in hosts with particularly low star formation rates. 

In practice, the mid- to far-infrared intrinsic AGN SEDs mentioned above, scaled according to the AGN luminosity, are often used to assess whether a given study is still in a regime where host dominance in the rest frame far-infrared can be confidently assumed or not. This is illustrated in Figure~\ref{fig:growth}, which shows a summary of average far-infrared luminosities (tracing host star formation rates) for AGN of different luminosities and redshifts. Pre-\herschel\ studies of samples of powerful AGN indicated correlations between SFR and AGN luminosity that are suggestive of merger-induced correlations \citep[e.g.][]{lutz08,netzer09}. The much better statistics and the breaking of redshift-luminosity degeneracies by the \herschel\ samples provides a more differentiated picture. Host SFRs at a given AGN luminosity rise steeply with redshift. At $z<0.8$, dependence of SFR on AGN luminosity is not significant for moderate luminosity AGN but appears to be present for the most luminous ones (bolometric $L_{AGN}\sim 10^{45}$\ergs\/). At $0.8<z<2.5$, no significant SFR variations are seen over several orders of magnitude in AGN luminosity. Typical star formation rates of optically selected QSOs at similar redshifts, as derived from stacking or maximum-likelihood analysis, also find a clear increase of host SFRs with redshift, and typically weak dependence on AGN luminosity \citep{serjeant10,bonfield11}. Powerful $z>2$ optically selected QSOs are found to be hosted on average in highly star forming systems \citep{serjeant10,mor12b,leipski13,netzer13}. This reinforces previous ground based (sub)mm studies of powerful high-z QSOs, by new measurements which are probing closer to the peak of the star forming SED component. The AGN luminosity range covered by these high-z QSOs is too small to probe for trends with luminosity in the same way as for X-ray AGN at $z\lesssim 2.5$, but some correlation of host star formation and AGN luminosity among very luminous AGN is suggested \citep[e.g.][]{netzer13}.

This absence or weakness of trends in average SFR with AGN luminosity among moderate luminosity ($L_{AGN}< 10^{45}$\ergs\/) X-ray selected AGN is reported by several studies with partly overlapping deep X-ray and \herschel\ datasets \citep{shao10,mullaney12a,rosario12,rovilos12}. The increase of AGN host SFRs with redshift (Figure~\ref{fig:growth}) is similar to the increase with redshift in SFR of main sequence star forming galaxies, a link that is strongly reinforced by studies that compare active and inactive galaxies of the same stellar mass and redshift. AGN host SFRs and SSFRs are enhanced with respect to mass matched galaxies if both star-forming and passive galaxies are included in the matched reference sample \citep[][]{santini12}, but are close to those of mass-matched star forming main-sequence galaxies only \citep{mullaney12a,santini12,rosario13a}. In other words, the hosts of moderate luminosity z$\lesssim$2 AGN resemble typical massive `main sequence' star forming galaxies at these redshifts, and are less likely to be quenched \citep{mullaney12a,rosario13a}. The similarity between AGN hosts and star forming galaxies of the same stellar mass extends to the distribution of rest frame $U-V$ colors \citep{rosario13a}. Because of dust, those colors are poor tracers of star formation, however. AGN populate the `green valley' of a color-mass diagram in a similar way as massive star forming galaxies, and far-infrared based SSFR locates many green valley AGN on the main sequence. Location on the green valley then does not correspond to ongoing quenching of star formation, and the green optical colors are indicative of dust reddening.

 The link between the hosts of moderate luminosity AGN and normal star forming galaxies provided by these \herschel\ studies is fully consistent with the lack of evidence for enhanced merger fractions, in morphological studies of AGN hosts at these redshifts \citep[e.g.][and references therein]{cisternas11,kocevski12}. It is also consistent with the absence of some correlations that are expected in at least some versions of the merger scenario: Obscured AGN as defined by either high X-ray obscuring column or by optical type 2 are not more star forming than unobscured AGN at the same redshift \citep{rosario12,rovilos12,merloni14}, there is no evidence for enhanced star formation in the hosts of those AGN accreting with the highest Eddington ratios \citep{rosario13b}, and HiBal and non-BAL QSOs do not differ significantly in their far-infrared properties \citep{caoorjales12}.

Focussing on more luminous $L_{2-10keV}>10^{44}$\ergs\ $1<z<3$ AGN, \citet{page12} reported reduced \herschel\ detection rates and a drop in average SFRs in comparison to moderate luminosity AGN. They interpret this in terms of star formation suppression by the AGN. This finding is in tension with the better statistics study of \citet[][Figure~\ref{fig:growth}]{rosario12} which finds no such drop, and the significant \herschel\ detection rates of luminous AGN at similar redshift in other studies \citep[e.g.][]{hatziminaoglou10,dai12}. It is also not reproduced in a larger statistics follow-up \citep{harrison12} with methodology closely matching that of \citet{page12}. The differences between these studies are probably caused by the combination of small number statistics and a considerable dispersion in individual SFRs that enter the average SFRs for a certain $L_{2-10keV}$, z bin. There likely is a fraction of hosts of luminous $z\sim 2$ AGN that are quenched (i.e. have very low SFR), even if the average host SFR of such luminous AGN is not lower than the average SFR for lower luminosity ones. 

Evidence for suppressed star formation has been reported in massive radio galaxies at $z<0.8$ and $L_K>1.5$L$_*$, in comparison to magnitude and color-matched inactive galaxies \citep{virdee13}. Similarly, radio excess AGN at $0.2\lesssim z\lesssim 3$ show lower specific star formation rates than X-ray AGN at similar redshifts \citep{delmoro13}. These two results may currently present the strongest \herschel\ based evidence for quenching in a subset of the high redshift AGN population. Of course, radio galaxies are not all passive. Strong star formation in some powerful high-z radio galaxies has been implied by previous ground-based submillimeter observations and is being studied with detailed \herschel\ SEDs \citep[e.g.][]{seymour11,barthel12,wylezalek13}.

Quenching by an AGN may also be responsible for the fast (up to $\sim 1000$~km/s)
massive (outflow rates up to $\sim 1200$~\msun\ yr$^{-1}$) molecular outflows found in local ULIRGs via both CO emission and \herschel\ OH absorptions \citep[e.g.][]{feruglio10,fischer10,sturm11,veilleux13,cicone13}. These could indicate a phase where the cold gas reservoir of these galaxies is expelled on short timescales within $\sim 10^6-10^8$~yr. 

\begin{figure}
\center
\includegraphics[width=6.cm]{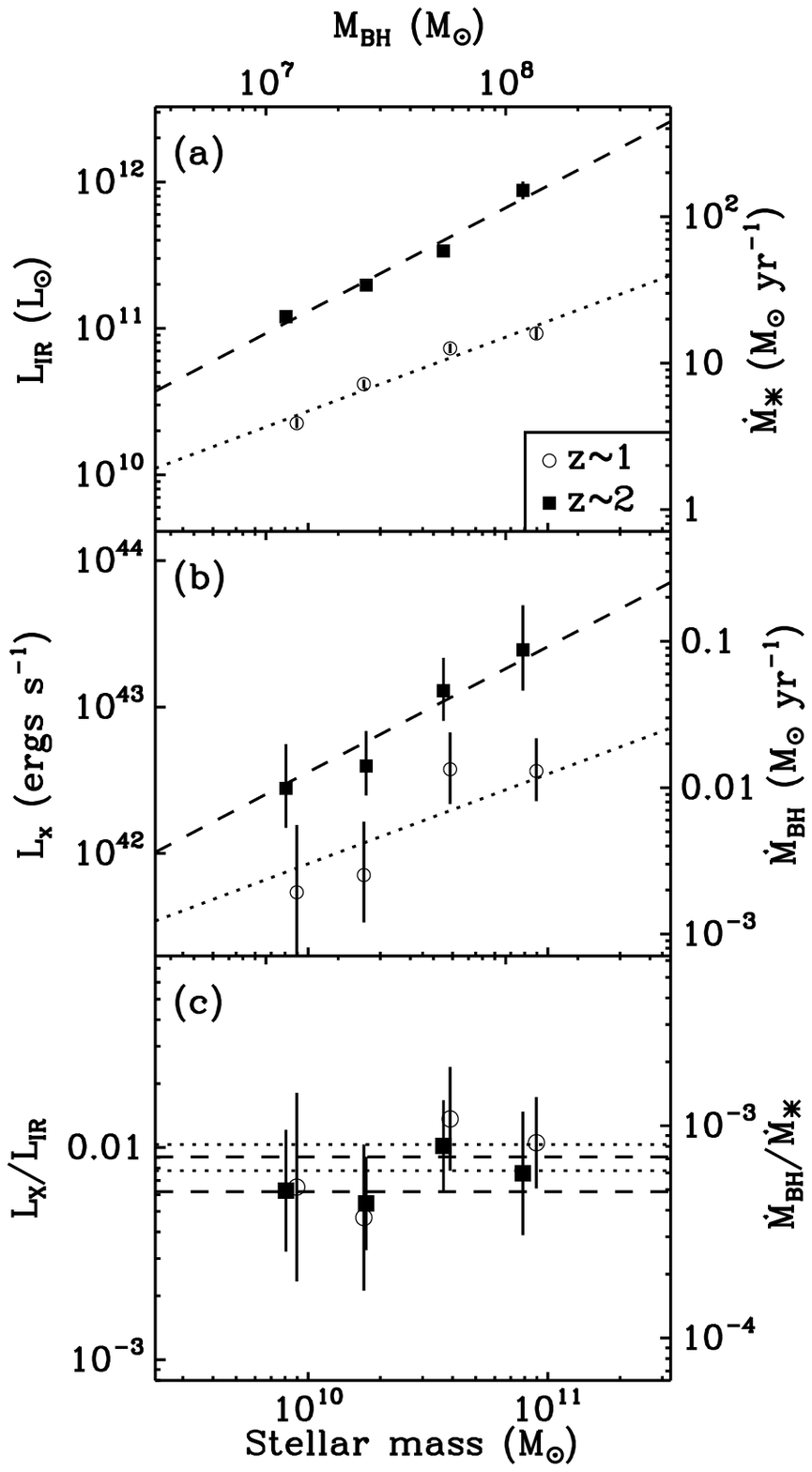}
\caption{(a) Average SFRs (right-hand axis) vs. stellar mass for $z\sim 1$
(open circles) and z$\sim 2$ (filled squares) samples of star forming galaxies (left-hand axis gives equivalent infrared luminosity for illustrative purposes only). Lines are least squares fits to the data. (b) Average X-ray luminosities of these star forming galaxies (same symbols as top panel) after accounting for any host galaxy contribution. Lines have the same gradients as in the top panel, only normalized to fit the inferred $\dot{M}_{\rm BH}$ which is indicated in the right-hand axis. (c) Average SMBH accretion rate to SFR ratio for the two redshift samples. The uncertainties on these points are consistent with a flat $\dot{M}_{\rm BH}$/SFR ratio with respect to \mstellar\ for both samples, indicated by the dotted and dashed lines, respectively. Reproduced from Figure~1 of \citet{mullaney12b}.} 
\label{fig:m12}
\end{figure}

The studies discussed above address the star formation rate of various AGN populations. Taking a reverse view, several recent works indicate a roughly proportional increase of mean black hole accretion rate with host star formation rate at $z<1$ \citep{rafferty11,chen13}. \citet{mullaney12b} group $z\sim 1$ and $z\sim 2$ star forming galaxies by stellar mass. Averaging over the samples in these stellar mass bins, they derive mean accretion and mean star formation rates which increase with host stellar mass at both redshifts. Absolute growth rate of the black hole and star formation is larger at z$\sim$2, but ratios of black hole growth and star formation are similar at both redshifts and all stellar masses (Figure~\ref{fig:m12}). In all three studies, the mean ratios of black hole to host growth are close to, but somewhat lower than the local ratio of black hole and bulge mass. A variety of factors may contribute to such mismatch: For example, obscured AGN may be missing from the samples studied. Also, it would be necessary to compare black hole accretion to only the fraction of the SFR that corresponds to bulge growth rather than disk growth. 

These results support an average tracking of star formation and black hole accretion in galaxy evolution since z$\sim$2. They plausibly fit the known trends of AGN fractions and star formation with stellar mass, and the similarity of the cosmic black hole accretion and star formation histories. It is important to note, however, that in all these studies, the use of sample means implies that black hole accretion is effectively averaged over the $\sim 10^{7\ldots 8}$yr timescales in which SFRs may vary, or over even longer `galaxy evolution' timescales. At any given time, most of the rapidly variable BH accretion actually happens in an AGN subset of the galaxies. While the mean relation of star formation and black hole accretion may be plausibly linked to gas that is supplied to both, the specific processes of feeding star formation and black hole on hugely different spatial scales remain little constrained by this tracking in the long term average.

Returning to Figure~\ref{fig:growth}, we conclude that the lack of correlation between SFR and AGN luminosity for $z<2.5$ moderate luminosity AGN, and the similarity of their hosts to main sequence star forming galaxies implies that they largely are part of the same population. The classification as AGN or as inactive as well as AGN luminosity are varying rapidly through processes that do not strongly relate to the global star formation rate. 

Merger induced effects may still play a role in shaping the increase in SFR with AGN luminosity for low-z and high luminosity AGN. This is in qualitative agreement with a visualisation of recent merger hydrodynamical models into this plane \citep{rosario12}, showing that these models spend considerable time in the region near the correlation line proposed by \citet{netzer09}. Fast AGN variability very likely plays a role in shaping this diagram, but different mechanisms deserve further scrutiny. Scenarios qualitatively reproducing Figure~\ref{fig:growth} range from one where accretion tracks constant star formation in the long term average but with a time-variable Schechter-like accretion rate distribution \citep{hickox13}, to one where accretion events and short star bursts mediated by (minor) mergers are placed on top of a slowly varying background star formation \citep{neistein14}.

In summary, \herschel\ studies suggest that z$\lesssim$2.5 AGN are hosted mostly in massive main sequence star forming galaxies with properties that are typical for their redshift. Special events such as major mergers likely play a less important role. Black hole feeding and star formation seem connected by the common gas reservoir and supply on a long galaxy evolution time scale. Given observational limitations such as the widespread use of averaging over subsamples, and the theoretical uncertainties, the detailed mechanisms of feeding the AGN are not strongly constrained.

\section{Dust as an interstellar medium tracer in distant galaxies}
\label{sect:dustism}

Measuring dust masses to study their scalings with other galaxy properties requires submm detections that in the past were typically available only for modest size samples, pre-selected either by the IRAS far-infrared emission or in the optical. This is now changing with unbiased large surveys and large local samples.

\citet{cortese12} use SPIRE data to study dust scaling relations for a magnitude- and volume limited sample of about 300 nearby galaxies, including star forming as well as passive objects. The ratio of dust mass to stellar mass decreases with stellar mass, stellar mass surface density, and NUV-r colour, and is significantly lower for early type galaxies. Such scalings are reminiscent of the equivalent scalings for the HI component \citep{catinella10} and molecular component \citep{saintonge11} of the interstellar medium in local galaxies. Dust stripping by environmental effects is indicated, but at a lower level than for HI, consistent with the fact that dust is tracing both extended atomic and more compact molecular material. \citet{bourne12} stack SPIRE data for a large $z<0.35$ optically selected spectroscopic sample. Again, dust masses increase with stellar mass while the $M_{dust}/M_*$ ratio decreases, and there are clear indications for an increase of dust masses with redshift in the redshift range studied. Evolution of the dust mass function is very prominent in the $z<0.5$ study of \citet{dunne11}. Dust masses increase by a factor $\sim$5 from local to $z\sim0.5$, and a similar increase is reported for the ratio of dust mass and stellar mass. \citet{santini13} bin the redshift -- stellar mass -- star formation rate space out to $z=2.5$ and investigate scalings between dust mass, stellar mass, and star formation rate from stacked \herschel\ data. Dust masses increase with stellar mass and with redshift and strongly correlate with SFR. There is less of a trend with redshift for dust mass at given SFR and stellar mass.

These observed dust scalings plausibly fit the context in which SFR and stellar mass are linked via the evolving star forming sequence, and the increase of gas fraction of normal galaxies with redshift \citep[e.g.][]{daddi10,tacconi10}.
Given rather short gas depletion times for high redshift galaxies, models of the cosmic evolution of dust will have to increasingly move from closed box assumptions to considering gas inflow, as well as outflow to and recycling from the intergalactic medium, all related to current equilibrium attempts to represent the interplay of gas flows and stellar mass buildup in galaxy evolution \citep[e.g.][]{lilly13}.
Another interesting question is to what extent it is observationally possible in this context to quantify the cold interstellar medium of high redshift galaxies via the dust mass rather than direct tracers of the gas.      

\subsection{Dust as a tracer of cold gas} 

The cold gas content of evolving high redshift galaxies is a key measurable for characterising the interplay of gas inflow, gas consumption by star formation, metal enrichment, and outflows or feedback, during steady evolution and while influenced by merging and other external effects. Traditionally, low lying rotational CO transitions are used to measure the molecular gas content. Even at the beginning of the ALMA era, $z>1$ CO detections remain hard to obtain, though, with of order 200 total detections \citep{carilli13} of which only about 60 refer to galaxies near the star-forming main sequence \citep{tacconi13}. In addition, use of CO as a molecular gas tracer relies on the CO (1-0) luminosity to molecular gas mass `conversion factor' which depends on physical conditions of the source and on metallicity \citep[see][for a recent review]{bolatto13}. Often, a step is involved of applying excitation corrections to measurements of CO lines originating in higher rotational states, rather than measuring the CO (1-0) for which the conversion to gas mass is calibrated.

\herschel\ has obtained a large number of dust continuum detections of high redshift galaxies, to which ALMA starts adding longer wavelength submm detections of unprecedented depth \citep[e.g.][]{scoville14}. These resources have spawned great interest in using dust as a proxy for cold gas content, applying locally calibrated conversions from observed rest frame far-infrared and submillimeter fluxes to dust mass, and a metallicity-dependent conversion from dust to cold gas mass. Several assumptions and steps enter such methods.

Conversions from dust emission SEDs to dust mass are calibrated in the Milky Way or nearby galaxies and implicitly assume dust properties similar to the ISM of these galaxies. ISM dust has a variety of sources including AGB star mass loss and supernova explosions. The relative weight of these can vary especially towards highest redshift, depending on the histories of star formation and outflow from the galaxy. In addition, dust destruction and re-formation in the ISM proper may play a significant role \citep[e.g.][]{jones96}, depending on physical conditions. Observational evidence against homogeneous dust properties includes long-known variations in the optical/UV extinction curves, variations of the submm emissivity index with dust temperature \citep[e.g.][note discussion in the literature on influence of fitting techniques]{paradis10}, presence of crystalline silicate dust in some local ULIRGs but not normal galaxies \citep{spoon06}, unusual silicate properties in some AGN spectra \citep{sturm05,markwick07}, and the rest frame $\gtrsim$500~$\mu$m submillimeter excess often appearing in the SEDs of low metallicity galaxies \citep[e.g.,][and references therein]{remyruyer13a}.   

Dust masses are most easily measured on the Rayleigh-Jeans tail of the SED where the dependence on dust temperature is only linear and there is almost no concern of optically thick emission. This wavelength condition is often not met for pure \herschel\ data, with the 500~$\mu$m end of the SPIRE range and in particular source confusion severely limiting the accessible rest wavelength range for high redshift galaxies, except for extremely luminous objects. High S/N color information is then needed instead to constrain either the dust temperature, or equivalent parameters such as radiation field intensity in physical dust emission models. In addition, such measurements will be insensitive to any very cold dust. \citet{draine07} demonstrate that for a local disk galaxy sample and the physical dust model of \citet{draineli07}, coverage out to rest wavelength 160~$\mu$m is sufficient to obtain dust masses with an extra error due to this wavelength constraint of less than a factor 2.2, and often less than a factor 1.5, for appropriately constrained model parameters. Similarly, \citet{dale12} compare dust masses of 61 local KINGFISH galaxies from fitting \citet{draineli07} models to either the \spitzer\ only data out to 160~\mum, or combined \spitzer\ and \herschel\ data out to 500~\mum. Masses from these two approaches agree well on average with only 0.2~dex dispersion, but the ratio has trends with both dust temperature and metallicity. Such use of $\lambda_{\rm rest}\leq 160\mu$m fits to derive dust masses may be transferable to high redshift galaxies, as long as the ISM conditions are similar to the local calibration sample and dust emission is not optically thick near the far-infrared peak.

Conversion from dust mass to (atomic plus molecular) gas mass $\delta_{\rm GDR}M_{\rm dust}=M_{\rm H_2}+M_{\rm HI}$ has to assume a metallicity dependence of the gas-to-dust ratio $\delta_{\rm GDR}$. This is typically assumed to be close to inverse linear \citep[e.g.][]{leroy11}, but steeper relations have also been argued for \citep{munozmateos09} or constancy assumed \citep{scoville12}. For a $z\sim 2$ lensed sample with measured metallicities, \herschel\ dust masses and CO measurements, \citet{saintonge13} deduce a relation of gas-to-dust ratio to metallicity that is larger by a factor 1.7 with respect to the local one of \citet{leroy11}. Irrespective of an origin in changed dust properties or changed CO conversion, this offset cautions about the transfer of local calibrations. Scatter around such relations is significant \citep{draineli07,remyruyer13b}, and a number of high redshift SMGs have been reported with apparently much smaller gas-to-dust ratio than expected from their nebular metallicity \citep{santini10}. Typically, individual metallicities are however unavailable for high-z galaxies, requiring to adopt scalings such as redshift dependent mass-metallicity relations \citep[e.g.][]{zahid13} or the fundamental metallicity relation of \citet{mannucci10} which links metallicity to stellar mass and star formation rate in a redshift-independent way. Differences between such scalings, and the largely unconstrained scatter of high redshift galaxies around the scaling can lead to uncertainties of a factor of a few in metallicity alone.  

These data practicalities and systematic effects suggest that dust based methods provide a largely independent second view on the cold gas content of high-z galaxies in addition to CO, but with significant uncertainty. 

For a sample of ten normal spiral galaxies in the local universe with spatially resolved high quality PACS and SPIRE photometry as well as CO and HI data, \citet{eales12} assumed homogeneous dust and gas properties and solved for minimal dispersion between dust based and CO+HI based estimates of the interstellar medium mass. They verified that this leads to a consistent picture with a CO conversion factor and dust emissivity close to commonly adopted values for the Milky Way.

At higher redshift, \citet{magdis11b} and \citet{magdis12a} study small samples
of galaxies with \herschel\ and (sub)mm detections and low-J CO data, applying the \citet{draineli07} dust model. Like \citet{magnelli12b}, they also find dust-based ISM masses in support of a decrease of the CO conversion factor $\alpha$ when moving from main sequence $z\sim 1-2$ galaxies to (merger) starbursts. Resorting to stacking because of the requirements that even the \citet{draineli07} model puts on photometric S/N and rest wavelength coverage, \citet{magdis12a} then study the properties of larger $z\lesssim 2.3$ samples near the main sequence. They find that SSFR trends across and along the main sequence are related mostly to varying gas fractions. Detected variations with redshift are a significant increase in gas fraction and a mild decrease in gas depletion times $M_{Gas}/SFR$, in agreement with CO based results \citep[e.g.][]{tacconi13}. \citet{santini13} use similar stacking methods for mostly near-main sequence bins in the redshift -- stellar mass -- star formation rate space out to $z=2.5$, finding a steeper decrease of gas depletion time and suggesting a fit by a redshift-independent fundamental relation between gas fraction, stellar mass, and star formation rate. 

Following a different approach, \citet{nordon13} derive a method for estimating the molecular gas content of near main sequence galaxies solely from more readily available integrated far-infrared luminosities and rest UV obscuration properties. Based on a study of the UV properties of Herschel galaxies and comparison to CO, they derive a quantity that encodes the attenuation contributed by the molecular gas mass per young star. These estimates thus include a next order correction, compared to the average gas depletion time of the calibration sample with CO data. \citet{berta13b} use this method as well as gas depletion time scalings from \citet{tacconi13} to derive the molecular gas mass function since z$\sim$3.

Dust emission has been successfully applied as an ISM tracer for high redshift galaxies, with results in satisfactory agreement with and partly calibrated on CO based studies. Moving from current samples to higher redshifts, lower stellar masses and lower metallicities will require more sensitive (and difficult) observations with ALMA, as well as larger extrapolation from the ISM conditions at which current dust based methods are calibrated.

\section{Outlook}

Productive mining of the current generation of far-infrared surveys will continue for years. Since infrared surveys are now firmly integrated into both science goals and 
field choice of multiwavelength surveys of galaxy evolution, their legacy value will only be strengthened as other facilities continue to expand our knowledge of the galaxy populations and survey fields under study. 

Fruitful links between \herschel\ and the current generation of (sub)mm interferometers such as the IRAM-PdBI indicate the unique role that such interferometers and in particular ALMA will assume in the detailed characterisation
of far-infrared detected galaxies and in strengthening our knowledge of their gas content, interstellar medium conditions, and kinematics.  

Concerning photometric surveys, large steps in the submillimeter regime can be expected by bringing improved detector technology to large telescopes at the best ground-based sites, such as planned by the CCAT project. In space, the planned Japanese SPICA large cryogenic space telescope will for the first time provide the far-infrared spectroscopic diagnostics which for high-z galaxies were effectively out of reach of the passively cooled \herschel\ telescope. With fast and deep mapping in the mid-infrared and in the far-infrared out to $\sim$200~\mum, SPICA will also open a new regime in size and quality of SED information for galaxy evolution studies.

\section{Summary points}
New observing capabilities from \herschel\ and other facilities have enhanced the power of far-infrared studies of galaxy evolution, for redshifts up to $z\sim 2$ and beyond.
\begin{enumerate}
\item Near its peak, three quarters of the cosmic infrared background is resolved into individually detected galaxies.
\item Far-infrared calorimetric star formation rates are now available for massive normal star forming galaxies at high redshift, and help calibrate other indicators. Direct far-infrared luminosity functions and star formation rate densities have been obtained out to $z\gtrsim 3$.
\item The interstellar medium conditions of high-z massive star forming galaxies, as expressed in the far-infrared SED and in the ratio of mid- and far-infrared emission,
are better described in relation to the evolving main sequence of star forming galaxies rather than by absolute infrared luminosity. Most star formation happens near the main sequence.
\item The far-infrared emission shows that AGN hosts out to $z\sim 2$ typically are normal massive star forming galaxies. The role of major mergers is less important.
\item Steps have been made towards using dust emission as a tracer of the total interstellar medium mass of high-z galaxies.
\end{enumerate}

\section{Disclosure statement}
The author is not aware of any affiliations, memberships, funding, or financial holdings that might be perceived as affecting the objectivity of this review.

\section{Acknowledgements}
The author would like to thank all who provided comments, insights, discussions, data, and figures, in particular  
Stefano Berta, 
David Elbaz,
Sandy Faber,
Reinhard Genzel,
Hagai Netzer, 
Benjamin Magnelli, 
Emma Rigby, 
David Rosario,
Stephen Serjeant,
Toshinobu Takagi,
  and Stijn Wuyts.

\bibliography{dlaraa}

\end{document}